\documentclass[11pt,a4paper,twoside,fleqn]{article}
\usepackage{color,graphicx,lscape,here,geometry,setspace,multirow,enumitem}
\usepackage[dvipsnames]{xcolor}
\usepackage{graphicx}
\usepackage{natbib}
\usepackage{amsfonts}
\usepackage{authblk}
\usepackage{silence}

\usepackage[misc,geometry]{ifsym}

\usepackage{rotating}
\usepackage{multirow}

\usepackage{booktabs}
\usepackage{diagbox}
\usepackage{color,colortbl}
\usepackage{setspace}
\usepackage{algorithm2e}
\usepackage{enumitem}
\usepackage{amsfonts,amsmath,amssymb}
\usepackage{tikz}

\definecolor{nblue}{HTML}{000660}
\usepackage[colorlinks=true,urlcolor=nblue,linkcolor=nblue,citecolor=nblue]{hyperref}
\usepackage{bigstrut}

\newcommand{\inlinehead}[1]{{\sffamily\noindent\textsc{\textbf{#1}}}. }

\usepackage{tabularx}
\usepackage{threeparttable}
\usepackage{dcolumn}
\newcolumntype{d}[1]{D{.}{.}{#1}}

\usepackage{array}
\newcolumntype{C}[1]{>{\centering\arraybackslash}p{#1}}

\usepackage{etoolbox} 
\makeatletter
\patchcmd{\BR@backref}{\newblock}{\newblock[}{}{}
\patchcmd{\BR@backref}{\par}{]\par}{}{}
\makeatother

\usepackage{pdflscape}
\usepackage{afterpage}

\usepackage{longtable}
\usepackage{array}

\usepackage{comment}
\usepackage{mathtools}
\usepackage[hang,flushmargin]{footmisc} 
\usepackage[hang]{footmisc}
\usepackage[british]{babel}

\pdfoutput=1
\usepackage{float}
\restylefloat{table}

\usepackage[title,titletoc]{appendix}
\makeatletter

\renewenvironment{appendices}{%
    \begin{oldappendices}%
    \renewcommand{\thefigure}{\ifnum \c@section>\z@ \thesection.\fi\@arabic\c@figure}%
    \@addtoreset{figure}{section}%
    \renewcommand{\thetable}{\ifnum \c@section>\z@ \thesection.\fi\@arabic\c@table}%
    \@addtoreset{table}{section}}{%
    \end{oldappendices}%
}\makeatother

\usepackage{titlesec} 
\titleformat{\section}[block]{\bfseries\sffamily\large}{\thesection. }{0em}{\MakeUppercase} 
\titleformat{\subsection}[block]{\bfseries\sffamily\large}{\thesubsection. }{0em}{} 
\titleformat{\subsubsection}[block]{\large}{}{0em}{\itshape} 

\let\natbibcitet\citet
\renewcommand\citet{\bibpunct{(}{)}{,}{a}{,}{,}\natbibcitet}
\let\natbibcitep\citep
\renewcommand\citep{\bibpunct{(}{)}{;}{a}{,}{;}\natbibcitep}
\newcommand{\bi}{\begin{itemize}}
\newcommand{\ei}{\end{itemize}}
\newcommand{\be}{\begin{equation}}
\newcommand{\ee}{\end{equation}}
\defcitealias{ieo14}{IEO, 2014}

\long\def\symbolfootnote[#1]#2{\begingroup%
\def\thefootnote{\fnsymbol{footnote}}\footnote[#1]{#2}\endgroup}

\makeatletter
\def\ubar#1{\underline{\sbox\tw@{$#1$}\dp\tw@\z@\box\tw@}}
\def\obar#1{\overline{\sbox\tw@{$#1$}\dp\tw@\z@\box\tw@}}
\makeatother

\usepackage{bm}
\usepackage{caption}
\usepackage{subcaption}

\makeatletter\let\p@subfigure\thefigure\makeatother

\floatstyle{plaintop}
\restylefloat{table}

\usepackage{fancyhdr} 
\usepackage{cleveref}
\crefname{chapter}{Chapter}{Chapters}
\crefname{section}{Section}{Sections}
\crefname{subsection}{Section}{Sections}
\crefname{subsubsection}{Section}{Sections}
\crefname{figure}{Figure}{Figures}
\crefname{table}{Table}{Tables}
\crefname{equation}{Equation}{Equations}
\crefname{appendix}{Appendix}{Appendices}
\crefname{appendices}{Appendix}{Appendices}

\crefname{appsec}{Appendix}{Appendices}

\makeatletter

\def\Autoref#1{%
  \begingroup
  \edef\reserved@a{\cpttrimspaces{#1}}%
  \ifcsndefTF{r@#1}{%
    \xaftercsname{\expandafter\testreftype\@fourthoffive}
      {r@\reserved@a}.\\{#1}%
  }{%
    \ref{#1}%
  }%
  \endgroup
}
\def\testreftype#1.#2\\#3{%
  \ifcsndefTF{#1autorefname}{%
    \def\reserved@a##1##2\@nil{%
      \uppercase{\def\ref@name{##1}}%
      \csn@edef{#1autorefname}{\ref@name##2}%
      \autoref{#3}%
    }%
    \reserved@a#1\@nil
  }{%
    \autoref{#3}%
  }%
}
\makeatother
\def\titletext{Nowcasting with Mixed Frequency Data \\ Using Gaussian Processes}

\title{\sffamily\huge{\textbf{\titletext}}}
\author{}
\date{}


\def\equationautorefname~#1\null{%
  Eq.~(#1)\null
}
\def\equationautorefname~#1\null{
Eq.~(#1)\null
}

\usepackage[utf8, latin1]{inputenc}                 

\begin{document}

\maketitle
\vspace*{-4.5em} 
\normalsize
\begin{center}
\begin{minipage}{.49\textwidth}
  \large\centering Niko \MakeUppercase{Hauzenberger}\\[0.25em]
  \footnotesize University of Strathclyde, United Kingdom
\end{minipage}
\begin{minipage}{.49\textwidth}
  \large\centering Massimiliano \MakeUppercase{Marcellino}\\[0.25em]
  \footnotesize Bocconi University, IGIER, Baffi, Bidsa, CEPR, Italy
\end{minipage}

\vspace*{1em}

\begin{minipage}{.49\textwidth}
  \large\centering Michael \MakeUppercase{Pfarrhofer}\\[0.25em]
  \footnotesize WU Vienna, Austria
\end{minipage}
\begin{minipage}{.49\textwidth}
  \large\centering Anna \MakeUppercase{Stelzer}\\[0.25em]
  \footnotesize Oesterreichische Nationalbank, Austria
\end{minipage}
\end{center}

\vspace*{1em}

\doublespacing
\begin{center}
\begin{minipage}{0.80\textwidth}
\noindent\small We develop Bayesian machine learning methods for mixed data sampling (MIDAS) regressions. This involves handling frequency mismatches and specifying functional relationships between many predictors and the dependent variable. We use Gaussian processes (GPs) and compress the input space with structured and unstructured MIDAS variants. This yields several versions of GP-MIDAS with distinct properties and implications, which we evaluate in short-horizon now- and forecasting exercises with both simulated data and data on quarterly US output growth and inflation in the GDP deflator. It turns out that our proposed framework leverages macroeconomic Big Data in a computationally efficient way and offers gains in predictive accuracy compared to other machine learning approaches along several dimensions.\\

{\sffamily\textbf{JEL}}: C11, C22, C53, E31, E37\\
{\sffamily\textbf{KEYWORDS}}: prediction, MIDAS, machine learning, Bayesian additive regression trees
\end{minipage}
\end{center}

\singlespacing\vfill\noindent{\footnotesize\textit{Contact}: Massimiliano Marcellino (\href{mailto:massimiliano.marcellino@unibocconi.it}{massimiliano.marcellino@unibocconi.it}), Department of Economics, Bocconi University. The views expressed in this paper are those of the authors and do not necessarily reflect the views of the Oesterreichische Nationalbank (OeNB) or the Eurosystem. We thank Eric Ghysels and participants at research seminars at the IHS, IIASA, OeNB, CIREQ-CMP Econometrics Conference and the IAAE 2024 for insightful comments and discussions. Hauzenberger and Pfarrhofer acknowledge funding by the Jubil\"aumsfonds of the OeNB, grants 18763, 18765. Codes and replication files are available at \href{https://github.com/mpfarrho/gp-midas}{github.com/mpfarrho/gp-midas}.}

\thispagestyle{empty}
\newpage\doublespacing\normalsize\renewcommand{\thepage}{\arabic{page}}
\renewcommand{\footnotelayout}{\setstretch{1.5}}

\section{Introduction}\label{sec:introduction}
This paper develops flexible nowcasting and short-horizon forecasting methods by combining elements from three strands of econometric literature. {First}, drawing from the mixed data sampling (MIDAS) framework proposed by \citet{ghysels2007midas} and \citet{andreou2010regression}, we leverage techniques that permit the efficient use of predictors sampled at a higher frequency than the target variable \citep[see][for a recent review]{ghysels2024econometric}. {Second}, the Big Data literature, which is based on the notion that exploiting a large set of predictors can improve predictive accuracy \citep[see, e.g.,][in the context of mixed frequency models]{babii2022machine,babii2024nowcasting,mogliani2021bayesian,mogliani2024bayesian,beyhum2023factor,borup2023mixed}. {Third}, the machine learning literature, which suggests that flexible models can improve predictive accuracy by uncovering complex relationships among variables (see, e.g., \citet{hasti2009elements} for a textbook, or \citet{goulet2022machine} in a macroeconomic context).

We use Gaussian Processes (GPs) to estimate the unknown and potentially nonlinear relationships between a target variable and a large set of mixed frequency predictors. When modeling the conditional mean of the MIDAS framework with a GP, we refer to this model as GP-MIDAS. GPs have been used previously in single frequency economic applications, see e.g., \citet{clark2024forecasting} in the context of inflation forecasting, or \citet{hauzenberger2024gaussian} who estimate effects of uncertainty shocks. By contrast, to handle mixed frequencies in this nonlinear context, we use several variants of restricted MIDAS polynomials in the spirit of \citet{ghysels2007midas}, or the unrestricted MIDAS (UMIDAS) approach by \citet{foroni2015unrestricted}, where each single high-frequency predictor is split into many low-frequency ones.

The restricted GP-MIDAS can be interpreted as a \textit{structured} (i.e., not \textit{randomly}) compressed GP \citep[see][for a discussion of the latter]{guhaniyogi2016compressed}, where the number of relevant predictors in our nonparametric model lies on a lower dimensional space. The MIDAS framework offers a natural angle to reduce the dimensionality of covariates. Therefore, it can improve the quality of feature extraction in a high-dimensional Big Data setting but also sharpen predictive inference \citep[see, e.g.,][]{snelson2012variable, paananen2019variable, liu2020gaussian}. Such a modeling strategy is therefore expected to be effective in disentangling signal from noise in a MIDAS regression, particularly when dealing with either noisy measurements of high-frequency predictors and/or substantial correlations among predictors (which are inherent due to the design of the MIDAS setup).

To assess its merits, we benchmark GP-MIDAS against other machine learning techniques that have recently gained popularity in the macroeconomic (mixed frequency) forecasting literature. As a nonparametric alternative to GPs for modeling the conditional expectation of the dependent variable in MIDAS models, we consider Bayesian Additive Regression Tree \citep[BART, originally proposed by][]{chipman2010bart} models. While GPs estimate nonlinear functions via a (theoretically infinite) mixture of Gaussian distributions, BART does so through a sum of regression trees. BART has been shown to perform quite well for nowcasting and forecasting with time series data in macroeconomic applications \citep[see, e.g.,][]{clark2023tail,clark2024investigating,huber2023nowcasting}. A recent review of GPs and BART in a (multivariate) time series context is provided in \citet{marcellino2024chapter}. 
Parametric machine learning regressions equipped with global-local shrinkage priors reflect the recent (Bayesian) MIDAS literature \citep[see, e.g.,][]{rodriguez2010mixed,mogliani2021bayesian,babii2022machine, kohns2023flexible}.\footnote{\citet{breitung2015forecasting} use small-scale homoskedastic nonparametric MIDAS models to predict inflation using daily data; \citet{chan2024tvpmidas} discuss how to efficiently introduce time-varying parameters in Bayesian MIDAS regressions. Other recent examples of how to use machine learning approaches in a MIDAS context are \citet{schnorrenberger2024harnessing} or \citet{arro2024nowcasting}.} This competitor is conditionally linear in parameters and thus allows us to gauge the role of considering nonlinearities of an unknown form. Besides, we implement all competing model specifications with and without stochastic volatility (SV) in the error terms, as SV has been useful for improving short-horizon density forecasts in previous work based on linear mixed frequency models \citep[see, e.g.,][]{carriero2015realtime,pettenuzzo2016midas,carriero2022nowcasting}. 

We develop efficient Bayesian estimation algorithms based on Markov chain Monte Carlo (MCMC) sampling, which permits the use of very large sets of predictors to produce point, density, and tail forecasts in a computationally efficient manner. After a careful examination of the relative empirical performance of the various methods with simulated data, we apply these models to now- and forecast quarterly annualized real GDP growth and inflation in the GDP deflator with monthly predictors in a large-scale pseudo real-time evaluation scheme. The competing model combinations differ in the size of the information set (with up to about $120$ predictors), their conditional mean and variance parameterizations, and several MIDAS weighting schemes. We consider a long holdout period starting in the early 1990s.

There are several findings to report. First, nonlinear means are relatively more important when trying to improve forecasts than nowcasts, although they remain competitive for nowcasting as well. GP outperforms BART among the nonparametric methods, and SV is an important model feature irrespective of other specification details. Second, the size of the information set matters slightly more for linear competitors; the nonparametric versions typically perform well already with small data input. Relatedly, unrestricted MIDAS versions are rarely among the best-performing specifications, and it is beneficial in most cases to structure the predictors with specific lag polynomials. Third, the nonparametric models are strongest when the least information is available. For instance, on average, they perform relatively better for somewhat longer-horizon predictions.

The rest of the paper is structured as follows. Section \ref{sec:econometrics} presents the econometric framework. Section \ref{sec:synthetic} assesses the performance of the methods with simulated data. Section \ref{sec:application} contains the empirical application on nowcasting and short-term forecasting of US GDP growth and inflation. Section \ref{sec:conclusions} summarizes the main findings and concludes. Additional details on the models, estimation and empirical results are provided in an Appendix.

\section{Econometric Framework}\label{sec:econometrics}
\subsection{A brief review of MIDAS}\label{subsec:review}
Let $\{y_{t}\}_{t=1}^{T_L}$ denote a scalar target variable on the lower of two or more frequencies, observed $T_L$ times; and $\{z_{t}\}_{t=1}^{T_H}$ is a single high-frequency predictor observed $T_H$ times. Assuming an evenly spaced frequency mismatch between the target and the predictor yields an integer ratio $m=T_H/T_L$; $m$ thus indicates how often the high-frequency variables are observed in terms of one observation of the low-frequency variable.\footnote{For instance, when linking a quarterly dependent variable to monthly predictors there are three months per quarter and we have $m=3$. In this case, the fractional time indexes $t-2/3$ and $t-1/3$ refer to the first and second month of quarter $t$, respectively. It is worth noting that our approach can easily be extended to feature uneven frequency mismatches and multiple frequencies. Each frequency could then be assigned its own lag order and mismatch ratio, which we avoid here to keep the notation simple.}

Our more general framework is presented in the next section. For illustration, we begin with a simple MIDAS regression model $y_t = \mathfrak{B}(\mathfrak{L}^{1/m},\tilde{\bm{b}}) z_t + \epsilon_t$ where $\epsilon_t$ is a zero-mean error term, $\mathfrak{B}(\mathfrak{L}^{1/m},\tilde{\bm{b}})$ is a weighting function indexed by the high-frequency lag operator, $\mathfrak{L}^{p/m}z_t = z_{t-p/m}$, and $\tilde{\bm{b}} = (\tilde{b}_0,\hdots,\tilde{b}_{\mathbb{L}})'$ is an $(\mathbb{L}+1)\times1$-vector of parameters. In the spirit of distributed lag models, the weighting function is parameterized, as a finite-dimensional approximation, with $\mathfrak{B}(\mathfrak{L}^{1/m},\tilde{\bm{b}}) = \sum_{p=0}^{P_H - 1} \mathrm{B}(p,\tilde{\bm{b}}) \mathfrak{L}^{p/m}$, where $\mathrm{B}(p,\tilde{\bm{b}}) = \sum_{l = 0}^{\mathbb{L}} \tilde{b}_{l}\varphi_{l}(p)$. The $\varphi_{l}(p)$'s are basis functions which we stack in $(\mathbb{L} + 1)$-vectors $\bm{\mathrm{w}}_p = (\varphi_0(p),\hdots,\varphi_{\mathbb{L}}(p))'$, with associated weights in $\tilde{\bm{b}}$. We may collect these in a $P_H \times (\mathbb{L}+1)$-matrix $\bm{W} = (\bm{\mathrm{w}}_0,\bm{\mathrm{w}}_1,\hdots,\bm{\mathrm{w}}_{P_H - 1})'$, such that the linear MIDAS regression can be written as:
\begin{equation}
y_t = \tilde{\bm{z}}_t'\bm{W} \tilde{\bm{b}} + \epsilon_t = \bm{x}_t'\tilde{\bm{b}} + \epsilon_t,\label{eq:MIDASlin}
\end{equation}
where $\tilde{\bm{z}}_t = (z_t, z_{t-1/m},\hdots,z_{t-(P_H-1)/m})'$ collects high-frequency lags of the predictor and $\bm{x}_t = \bm{W}'\tilde{\bm{z}}_t$. Define the $P_H$-vector $\bm{b} = \bm{W}\tilde{\bm{b}}$, then it is easy to see that this imposes specific shapes on the implied lag-specific parameters $\bm{b} = (b_{0},\hdots,b_{P_H-1})'$, that are weighted averages of the basis functions: $b_p = \sum_{l=0}^{\mathbb{L}} \tilde{b}_l \varphi_{l}(p)$ for $p = 0,\hdots,P_H-1$.

Loosely speaking, the matrix of weights maps the $P_H$-sized predictor vector to a (usually) lower dimension $\mathbb{L} + 1$, thereby reducing the number of parameters that need to be estimated at the cost of some lost flexibility. We break the linearity in the relationship between ${y}_t$ and $\bm{x}_t$ imposed in Eq. (\ref{eq:MIDASlin}) by assuming:
\begin{equation*}
    y_t = f(\bm{x}_t) + \epsilon_t = f(\bm{W}'\tilde{\bm{z}}_t) + \epsilon_t,
\end{equation*}
where $f(\bullet)$ is an unknown function that links a low-dimensional aggregate of the high-frequency lags nonlinearly to the target variable. On the one hand, we thus exploit the MIDAS setup with distinct basis functions to reduce the dimensionality of the problem in a structured way (i.e., subject to sensible parametric restrictions). On the other hand, we regain flexibility (which was lost due to finite-dimensional approximations and choices about basis functions) by estimating a possibly nonlinear function instead of a deterministic functional form. Rather than a linear combination of basis functions we obtain a nonlinear one. In our favored implementation, we rely on a GP prior to infer this function from the data.

The MIDAS structure (and underlying weighing approach) has several interesting and potentially favorable implications for common kernels used with GPs. It also offers computational advantages. Before we discuss these in detail, we lay out our general framework, which allows for many high-frequency predictors.

\subsection{The general framework}\label{sec:econometrics_midas}
Let $\{y_{t}\}_{t=1}^{T_L}$ again be the scalar target variable and $\{\bm{x}_t\}_{t=1}^{T_L}$ now denotes an $M\times1$-vector comprised of predictors on the lowest of two or more frequencies, observed $T_L$ times. We consider regressions of the form:
\begin{equation}
    y_{t} = f(\bm{x}_{t}) + \epsilon_{t}, \quad \epsilon_{t}\sim\mathcal{N}(0,\sigma^2_{t}).\label{eq:midas}
\end{equation}
The (potentially unknown and nonlinear) conditional mean function $f:\mathbb{R}^M\rightarrow\mathbb{R}$ will be specified below, and $\epsilon_{t}$ is a zero mean conditionally Gaussian error term with variance $\sigma^2_{t}$. As discussed in the single-predictor case above, the mixed frequency aspects of our work are due to the way the vector $\bm{x}_t$ is constructed. The goal is to model the low-frequency variable as a function of high-frequency variables. Let $\{\bm{z}_t\}_{t=1}^{T_H}$ with $\bm{z}_t = (z_{1t},\hdots,z_{Kt})'$ denote a $K \times 1$-vector of variables observed $T_H$ times on a higher frequency. Further, $\tilde{\bm{z}}_{kt} = (z_{kt},z_{kt-1/m},z_{kt-2/m},\hdots,z_{kt-(P_H-1)/m})'$ is a $P_H\times1$-vector of high-frequency lags of the $k$th predictor.

In our empirical work, the vector $\bm{x}_t$ contains $P_L$ lags of the dependent variable and $P_H$ lags of the $K$ predictors:
\begin{equation*}
\bm{x}_t = (y_{t-1},\hdots,y_{t-P_L},\tilde{\bm{z}}_{1t}'\bm{W}, \hdots, \tilde{\bm{z}}_{Kt}'\bm{W})',
\end{equation*}
where $\bm{W}$ is the $P_H \times (\mathbb{L}+1)$ matrix of weights determined by the basis functions used to compress the high-frequency lags. Effectively, we thus have $M = P_L + K(\mathbb{L} + 1)$ predictors. We note that $\bm{x}_t$ could be augmented to include deterministic terms, lags of other low-frequency variables and latent or observed factors. 

\inlinehead{Restricted and Unrestricted MIDAS} When using UMIDAS (abbreviated \texttt{u} later) we have $\mathbb{L} + 1 = P_H$ and $\bm{W} = \bm{I}_{P_H}$ where $\bm{I}_N$ refers to an $N$-dimensional identity matrix. Hence, for each additional high-frequency indicator, UMIDAS results in $P_H$ additional covariates in Eq. (\ref{eq:midas}), leading to a rapid increase in the number of parameters, since $M = P_L + K P_H$. For an overview and a more detailed discussion, see \citet{foroni2015unrestricted}. It suffices to note here that UMIDAS neither imposes restrictions nor explicitly models the distributed lag structure of the predictor-specific high-frequency lags.

Restricted MIDAS specifications alleviate overparameterization concerns via setting $\mathbb{L} \leq P_H$. Several parsimonious parameterizations of the weight function are available. We briefly summarize the main approaches that will serve as competing variants in our empirical work. In many cases, it is convenient to impose specific shapes on the weights and thus emphasize distinct high-frequency lags explicitly --- for instance, to reflect that more recent lags are usually more important than more distant ones. This is the first option we consider, following \citet{ghysels2007midas}, referred to as the exponential Almon lag (\texttt{xalm}). Here, $\bm{W}$ collapses to a $P_H\times1$-vector and an implementation with two parameters sets its $(r+1)$th element to:
\begin{equation*}
\bm{W}_{[r+1]} = \frac{\exp(\theta_1 r) + \exp(\theta_2 r^2)}{\sum_{\tilde{r}=0}^{P_H-1} \exp(\theta_1 \tilde{r}) + \exp(\theta_2 \tilde{r}^2)},
\end{equation*}
for $r = 0,\hdots,P_H-1$. While allowing for a nonlinear weighing function, we consider this implementation as an $\mathbb{L} = 1$ case due to $\bm{W}$ being a vector. Estimating $\theta_1,\theta_2$ allows for straightforward high-frequency lag selection in a data-driven manner. In particular, the shape of the weight function translates immediately into the implied lag polynomial on the coefficients (when assuming linearity). An important special case arises for $\theta_1 = \theta_2 = 0$ which yields equal weights. This is commonly referred to as a bridge model (\texttt{br} later on), and represents the simplest case we consider in our empirical work.

More generally, ``indirect'' parameterizations of the weights may be superior because they offer additional flexibility. We refer to these cases as $\mathbb{L} > 1$. Among these are non-orthogonalized power polynomials, usually referred to as \citet{almon1965distributed} polynomials of degree $\mathbb{L}$, abbreviated as \texttt{alm}. These can be implemented by setting $\bm{\mathrm{w}}_p = (1,p,p^2,\hdots,p^{\mathbb{L}})'$, i.e., the associated basis functions are $\varphi_l(p) = p^l$. However, orthogonal polynomials are often preferable due to the reduced multicollinearity of compressed predictors. \citet{babii2022machine}, for instance, suggest to use Legendre polynomials (\texttt{leg}) shifted to the interval $[0,1]$. \citet{mogliani2024bayesian} propose to use Bernstein polynomials (\texttt{ber}), while \citet{chan2024tvpmidas} discuss using Fourier basis functions (\texttt{fou}). We implement several of these options and provide additional details in Appendix \ref{app:technical}.

The choices about $\bm{W}$ and $\mathbb{L}$ (i.e., how to compress the predictors in an explicitly structured way) have several compelling implications when used in conjunction with GPs. We introduce GPs and discuss what these implications are next.

\subsection{Gaussian processes for the conditional mean}\label{sec:econometrics_mean}
Rather than assuming a specific functional form we treat the conditional mean function $f(\bm{x}_{t})$ as unknown in our most general implementations and impose a GP prior directly on the functional relationship \citep[see][]{williams2006gaussian}:
\begin{equation}
    f(\bm{x}_{t}) \sim \mathcal{GP}\left(0,\mathcal{K}_{\bm{\kappa}}(\bm{x}_{t},\bm{x}_{t})\right),\label{eq:GPsingle}
\end{equation}
where $\mathcal{K}_{\bm{\kappa}}(\bullet)$ denotes a suitable covariance function (the kernel) which we define below. The properties of the conditional mean function, such as stationarity or smoothness, can be defined with distinct functional forms subject to a moderate number of tuning parameters. These are encoded in the vector $\bm{\kappa}$. Stacking the function values and data for a finite sample of $T_L$ observations, we obtain $\bm{f} = (f(\bm{x}_{1}),\hdots,f(\bm{x}_{T_L}))'$ and $\bm{X} = (\bm{x}_{1},\hdots,\bm{x}_{T_L})'$. The Gaussian process prior then takes the form of a $T_L$-dimensional multivariate Gaussian distribution:
\begin{equation}
    \bm{f} \sim \mathcal{N}\left(\bm{0}_{T_L},\mathcal{K}_{\bm{\kappa}}(\bm{X},\bm{X})\right),\label{eq:GPmulti}
\end{equation}
where we assume zero means, and the $(t,\tilde{t})$th element of $\mathcal{K}_{\bm{\kappa}}(\bm{X},\bm{X})$ is given by $\mathcal{K}_{\bm{\kappa}}(\bm{x}_t,\bm{x}_{\tilde{t}})$. Put simply, the value of the covariance function governs the prior association of the functional values between time $t$ and $\tilde{t}$ conditional on the input vectors $\bm{x}_t$ and $\bm{x}_{\tilde{t}}$.

\inlinehead{Kernel} It is common to define kernels based on a distance metric and a few hyperparameters. We rely on the Euclidean distance and use a squared exponential kernel:
\begin{equation}
    \underbrace{\mathcal{K}_{\bm{\kappa}}(\bm{x}_t,\bm{x}_{\tilde{t}})}_{\text{Cov}\left(f(\bm{x}_t),f(\bm{x}_{\tilde{t}})\right)} = \xi \cdot \exp \left( -\frac{1}{2} (\bm{x}_t - \bm{x}_{\tilde{t}})'\bm{\Lambda}(\bm{x}_t - \bm{x}_{\tilde{t}}) \right),\label{eq:kernel}
\end{equation}
where $\bm{\Lambda}=\text{diag}(\lambda_1,\hdots,\lambda_M)$. A few general comments about the hyperparameters are in order. The unconditional variance of the prior, $\xi$, is sometimes also referred to as the \textit{signal} variance (as opposed to the \textit{noise} variance $\sigma_t^2$). This labeling can best be justified by noting that the marginal variance of the observations is Var$(y_t) = \xi + \sigma_t^2$. Notice also that $\lim_{(\bm{x}_t - \bm{x}_{\tilde{t}}) \rightarrow \infty}\mathcal{K}_{\bm{\kappa}}(\bm{x}_t,\bm{x}_{\tilde{t}}) = 0$; that is, the more the values of the predictors differ between $t$ and $\tilde{t}$, the less likely there is any noteworthy association in the function values a priori. 

The so-called inverse length scales $\lambda_i$ in $\bm{\Lambda}$, by contrast, govern how quickly the conditional mean function varies with respect to changes in the $i$th input. As $\lambda_i\rightarrow0$, the impact of the $i$th predictor vanishes. Mechanically, this is due to the exclusion of $x_{it}$ from the distance metric, and $\lambda_i$ thus provides a rough gauge of variable importance. In practice, having predictor-specific inverse length scales is computationally costly and unreliable \citep[see][for a recent discussion]{dance2022fast}. Usually one thus assumes a common inverse length scale, such that $\bm{\Lambda} = \lambda\bm{I}_M$, which reduces the number of hyperparameters in $\bm{\kappa} = (\xi,\lambda)'$, see also \citet{guhaniyogi2016compressed} for a recent example.\footnote{We have experimented with alternative kernel parameterizations, such as using a Mat\'ern covariance function, but found the squared exponential kernel to perform very similarly, albeit with less hyperparameters to estimate.}

\inlinehead{Inference} Bayesian inference is based on manipulating the Gaussian distribution given by Eq. (\ref{eq:GPmulti}). Using stacked notation such that $\bm{y} = (y_1,\hdots,y_{T_L})'$ and $\bm{\epsilon} = (\epsilon_1,\hdots,\epsilon_{T_L})'$, we obtain $\bm{y}\sim\mathcal{N}(\bm{0},\mathcal{K}_{\bm{\kappa}}(\bm{X},\bm{X}) + \bm{\Sigma})$ and $\bm{\Sigma} = \text{diag}(\sigma_{1}^2,\hdots,\sigma_{T_L}^2)$. In the machine learning jargon, $\bm{X} \in \mathbb{R}^{T_L\times M}$ is usually referred to as ``training'' inputs. For a general discussion, let $\tilde{\bm{X}} \in \mathbb{R}^{\tilde{T}_L\times M}$ denote the corresponding ``test'' values for which we would like to obtain inference about the unknown function values \citep[see][for details]{williams2006gaussian}. This can be achieved by considering their joint distribution:
\begin{equation*}
    \begin{bmatrix}
        \bm{y}\\
        \tilde{\bm{f}}
    \end{bmatrix}\sim\mathcal{N}\left(\bm{0},
    \begin{bmatrix}
        \mathcal{K}_{\bm{\kappa}}(\bm{X},\bm{X}) + \bm{\Sigma}) & \mathcal{K}_{\bm{\kappa}}(\bm{X},\tilde{\bm{X}})\\
        \mathcal{K}_{\bm{\kappa}}(\tilde{\bm{X}},\bm{X}) & \mathcal{K}_{\bm{\kappa}}(\tilde{\bm{X}},\tilde{\bm{X}})
    \end{bmatrix}
    \right),
\end{equation*}
where $\tilde{\bm{f}}$ denotes the conditional mean function at the test inputs. Exploiting the properties of the multivariate Gaussian, the posterior is yet another multivariate Gaussian, with moments:
\begin{align}
    \mathbb{E}\left(\tilde{\bm{f}}|\bm{y}\right) &= \mathcal{K}_{\bm{\kappa}}(\tilde{\bm{X}},\bm{X}) \left[\mathcal{K}_{\bm{\kappa}}(\bm{X},\bm{X}) + \bm{\Sigma}\right]^{-1}{\bm{y}},\label{eq:gppost}\\
    \text{Var}\left(\tilde{\bm{f}}|\bm{y}\right) &= \mathcal{K}_{\bm{\kappa}}(\tilde{\bm{X}},\tilde{\bm{X}}) - \mathcal{K}_{\bm{\kappa}}(\tilde{\bm{X}},\bm{X})\left[\mathcal{K}_{\bm{\kappa}}(\bm{X},\bm{X}) + \bm{\Sigma}\right]^{-1} \mathcal{K}_{\bm{\kappa}}(\bm{X},\tilde{\bm{X}}).\nonumber
\end{align}

Computationally, this implies operations with matrices that are at most $\overline{T}_L = \max(T_L,\tilde{T}_L)$ dimensional. Specifically, the computationally costly operations are a matrix inversion, and, when it is required to sample from this distribution, a Cholesky decomposition. The computational complexities in general are thus cubic in $\overline{T}_L$ \citep[see, e.g.,][for a discussion of the scalability of GPs]{banerjee2013efficient}. The number of predictors $M$, however, is by and large unimportant from a computational viewpoint. 

This makes the MIDAS case particularly attractive for applying GPs because, on the one hand, we work with the lowest of all available frequencies, which means the least amount of observations. The side effect of increasing the number of predictors (particularly in the case of UMIDAS), on the other hand, and different from the case when assuming linearity, does not affect our framework other than implying an alternative covariance structure as captured with the kernel.

\subsection{Implications of MIDAS weighting for GPs}
Another way of writing the GP is to consider its so-called weight-space representation. The prior in Eq. (\ref{eq:GPmulti}) may be written as a conditionally linear regression:
\begin{equation*}
    \bm{y} = \bm{\Psi}(\bm{X}) \bm{\eta} + \bm{\epsilon}, \quad \bm{\eta}\sim\mathcal{N}\left(\bm{0},\bm{I}_{T_L}\right), \quad \bm{\epsilon}\sim\mathcal{N}(\bm{0},\bm{\Sigma}), \quad \bm{\Psi}(\bm{X})\bm{\Psi}(\bm{X})' = \mathcal{K}_{\bm{\kappa}}(\bm{X},\bm{X}).
\end{equation*}
It can be shown that this corresponds to a Bayesian linear regression with infinitely many basis functions \citep[see chapter 4.3 of][for details]{williams2006gaussian}. For an observation at time $t$ we have:
\begin{equation*}
    y_t = \bm{\psi}_t(\bm{X})'\bm{\eta} + \epsilon_t, \quad y_t|\bullet \sim \mathcal{N}\left(\sum_{\mathrm{t}=1}^{T_L} \eta_{\mathrm{t}} \cdot \psi_{t\mathrm{t}}(\bm{X}),\sigma_t^2\right)
\end{equation*}
where $\bm{\psi}_t(\bm{X})'$ represents the $t$th row of $\bm{\Psi}(\bm{X})$ and $\psi_{t\mathrm{t}}(\bm{X})$ is the $\mathrm{t}$th element of this vector. This representation illustrates that our approach shares similarities with \citet{mogliani2024bayesian}; in our case, the conditional distribution of $y_t$ is defined in terms of all observations --- due to our approach of modeling nonlinearities in the conditional mean --- rather than groups of predictors that arise from MIDAS-related schemes.

It is also worthwhile to assess a closely related alternative way of writing the posterior in Eq. (\ref{eq:gppost}); using $\bm{\mathfrak{w}} = (\mathfrak{w}_1,\hdots,\mathfrak{w}_{T_L})' = \left[\mathcal{K}_{\bm{\kappa}}(\bm{X},\bm{X}) + \bm{\Sigma}\right]^{-1}{\bm{y}}$, we obtain $\mathbb{E}(\tilde{\bm{f}}|\bm{y}) = \sum_{t=1}^{T_L} \mathfrak{w}_t \mathcal{K}_{\bm{\kappa}}(\tilde{\bm{X}},\bm{x}_t)$. Let us revisit the single high-frequency predictor case of Section \ref{subsec:review} and consider the in-sample functional values where $\tilde{\bm{Z}} = (\tilde{\bm{z}}_1,\hdots,\tilde{\bm{z}}_{T_L})'$ such that $\bm{X} = \tilde{\bm{Z}}\bm{W}$:
\begin{equation*}
   \mathbb{E}\left({\bm{f}}|\bm{y}\right) = \sum_{t=1}^{T_L} \mathfrak{w}_t \mathcal{K}_{\bm{\kappa}}\left(\tilde{\bm{Z}}\bm{W}, \bm{W}'\tilde{\bm{z}}_t\right), \quad \bm{\mathfrak{w}} = (\mathfrak{w}_1,\hdots,\mathfrak{w}_{T_L})' = \left[\mathcal{K}_{\bm{\kappa}}(\tilde{\bm{Z}}\bm{W}, \tilde{\bm{Z}}\bm{W}) + \bm{\Sigma}\right]^{-1}{\bm{y}}.
\end{equation*}
These alternative expressions illustrate two main aspects. First, Eq. (\ref{eq:GPsingle}) defines a theoretically infinite dimensional set of basis functions; conditioning on the actual sample yields a finite-dimensional representation. Second, the functional values are a linear combination of the kernel functions (which act as basis functions). These are in turn shaped by the presence of the MIDAS-related basis functions encoded in $\bm{W}$, i.e., how the high-frequency predictors are compressed.

This structure relates to the literature on compressed GP regression \citep[see][]{snelson2012variable, guhaniyogi2016compressed, paananen2019variable, liu2020gaussian}. The key assumption for compressed nonparametric regressions is that the relevant predictors can be projected into a lower dimensional feature space. In a MIDAS framework, the compression of the feature and/or predictor space arises from taking into account the mixed frequency nature of the target variable $y_t$ and the raw predictors $\tilde{\bm{z}}_t$. Specifically, the weighting scheme implies variation in length scales across high-frequency predictors.

To investigate these claims more formally, it is instructive to consider the kernel written explicitly in terms of the high-frequency lags. To simplify notation, we ignore the lags of the dependent variable for the moment and let $\tilde{\bm{z}}_t = (\tilde{\bm{z}}_{1t}',\hdots,\tilde{\bm{z}}_{Kt}')'$, such that $\bm{x}_t = (\bm{I}_K \otimes \bm{W}')\tilde{\bm{z}}_t$. We may then write the implied kernel as:
\begin{equation*}
    \mathcal{K}_{\bm{\kappa}}(\bm{x}_t,\bm{x}_{\tilde{t}}) = \xi \cdot \exp \left( -\frac{\lambda}{2} (\tilde{\bm{z}}_t - \tilde{\bm{z}}_{\tilde{t}})'\underbrace{(\bm{I}_K \otimes \bm{W})(\bm{I}_K \otimes \bm{W}')}_{\tilde{\bm{\Lambda}}}(\tilde{\bm{z}}_t - \tilde{\bm{z}}_{\tilde{t}}) \right).
\end{equation*}
Notice that this corresponds to Eq. (\ref{eq:kernel}), which results from $\bm{\Lambda} = \lambda \tilde{\bm{\Lambda}}$, and thus the weights can be interpreted as emphasizing or de-emphasizing which lags enter the distance metric in the covariance function. 

The implied inverse length scale matrix $\bm{\Lambda} = \bm{I}_K \otimes (\lambda\bm{W}\bm{W}')$ is block diagonal, and each of the blocks is associated with the high-frequency lags grouped by individual predictor. The common inverse length scale, $\lambda$, governs the ``global'' level of variability of the conditional mean function, and $\bm{W}$ provides distinct ``local'' adjustments through the MIDAS lags. This yields a family of ``MIDAS kernels'' among the class of squared exponential kernels, each with different implications for the covariance structure of the low-frequency functional values.

\begin{figure}[t]
    \includegraphics[width=\textwidth]{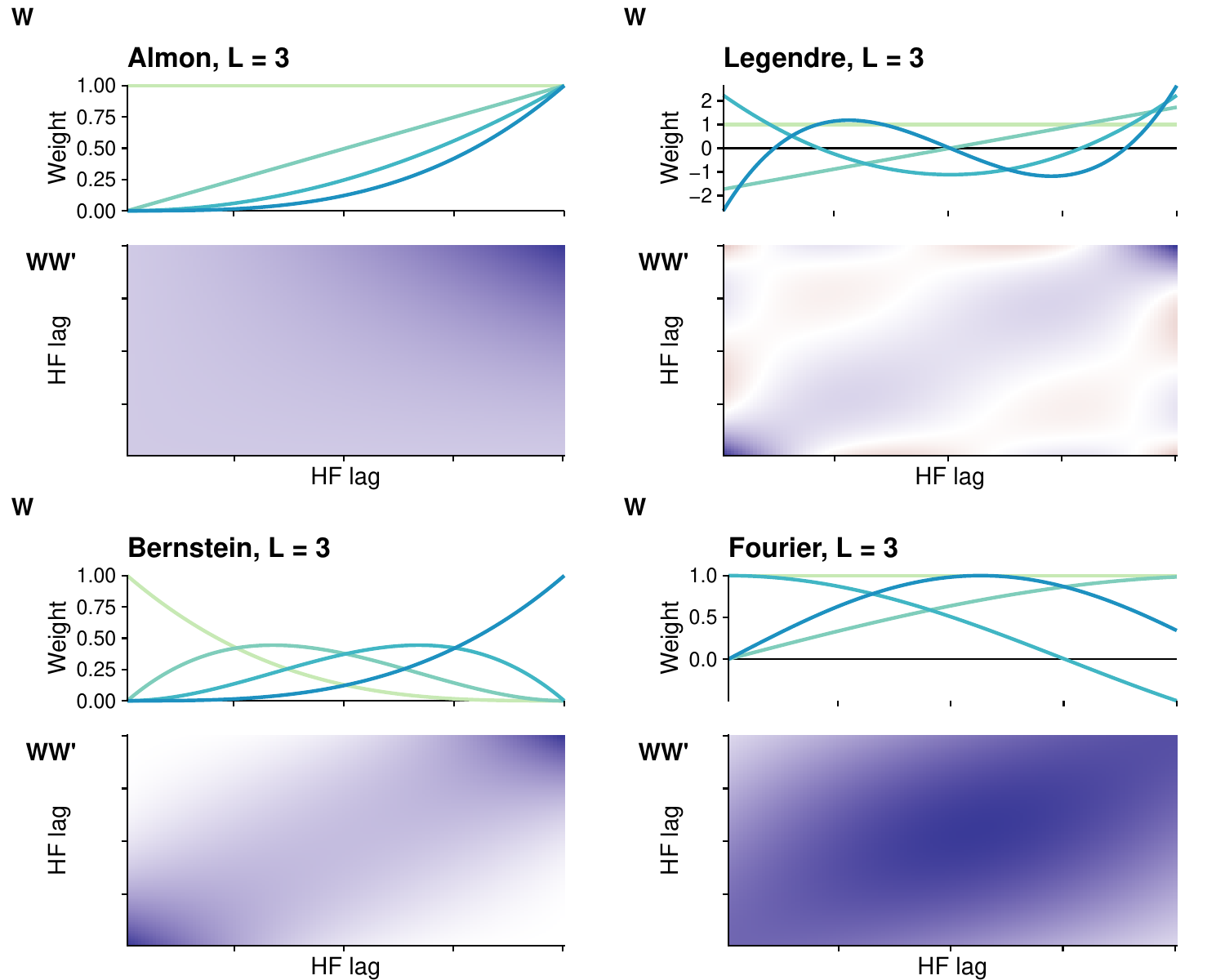}
    \caption{Implied inverse length scale matrix $\bm{\Lambda} = \lambda\bm{W}\bm{W}'$ and lag polynomials for $\lambda = 1$ and polynomial degree $\mathbb{L}=3$. Colors in the upper panels group the polynomials by degree and intensifying shades of blue indicate larger values in the lower panels.}\label{fig:kernelcrossprod}
\end{figure}

Figure \ref{fig:kernelcrossprod} provides a visual representation of the weights associated with each of the high-frequency lags, and the resulting implied matrix of inverse length scales per predictor, $\bm{\Lambda} = \lambda \bm{W}\bm{W}'$. We show only the $\mathbb{L}>1$ versions except UMIDAS which does not impose any structure and the resulting $\bm{\Lambda}$ is an identity matrix; the \texttt{br} implementation yields a matrix whose values are all equal to $P_H^{-2}$ and \texttt{xalm} specifies the shape of the weights explicitly. The \texttt{ber} and \texttt{fou} polynomials result in rather similar patterns, the former being more sparse; \texttt{alm} by comparison has more one-sided emphasis, whereas \texttt{leg} exhibits the most complex pattern across HF-lags.

\subsection{Alternative conditional mean specifications}
We argued above that the GP prior is well-suited for the MIDAS framework, because of the way it handles variables and the rather general form of nonlinearities it can capture. There are, however, alternative ways of how to estimate the function $f(\bm{x}_{t})$. We briefly discuss the main features of such alternatives below and provide further details in Appendix \ref{app:technical}.

The first option we consider are regression trees. In our empirical work, we consider a Bayesian implementation as a competing specification and rely on BART:
\begin{equation*}
    f(\bm{x}_{t}) \approx \sum_{s=1}^S \ell_{s}(\bm{x}_{t}|\mathcal{T}_s,\bm{\mu}_s),
\end{equation*}
where $\mathcal{T}_s$ denotes the tree structure, $\bm{\mu}_s$ is a vector of terminal node parameters associated with each tree, and we have $s=1,\hdots,S,$ regression trees. For these, we use the default prior setup and tuning parameters suggested in \citet{chipman2010bart}, which have been shown to work well also in a time series context \citep[see, e.g.,][]{clark2023tail}.

Another, and arguably our simplest variation (see also Section \ref{subsec:review}), is the linear MIDAS regression:
\begin{equation*}
    f(\bm{x}_{t}) = \bm{x}_{t}'\bm{\beta},
\end{equation*}
where $\bm{\beta}$ is an $M \times 1$ vector of regression coefficients. This represents the basic framework used in recent related papers that are inspired by the machine learning literature. \citet{babii2022machine} serves as an example of a classical econometric implementation, while Bayesian versions are developed in \citet{mogliani2021bayesian,mogliani2024bayesian}. Methodologically, these papers use variants of grouped shrinkage on the regression parameters, where groups are comprised of the compressed high-frequency lags for each of the $K$ predictors. Another noteworthy example is \citet{kohns2023flexible}, who add a time-varying intercept term and $t$-distributed errors with stochastic volatility. 

Our approach to inference is closely related to these implementations of (penalized or regularized) linear MIDAS regressions. We impose global-local shrinkage on the parameters $\bm{\beta}$ with a horseshoe prior \citep[HS,][]{carvalho2010horseshoe}. Below we refer to this specification as Bayesian linear regression (BLR). 

\subsection{Other specification details}\label{sec:econometrics_variance}
\inlinehead{Error Variance} For the case of homoskedastic errors (labeled \texttt{hom}), we assume $\sigma_{t}^2 = \sigma^2$ and $\sigma^2\sim\mathcal{G}^{-1}(a_0,b_0)$. To introduce stochastic volatility (SV), similar to \citet{pettenuzzo2016midas} but without predictors in the state equation, let $\varsigma_{t} = \log(\sigma_{t}^2)$ and assume:
\begin{equation}
    \varsigma_{t} = \mu_\varsigma + \phi_\varsigma (\varsigma_{t-1} - \mu_\varsigma) + \sigma_\varsigma\zeta_{t}, \quad \zeta_{t} \sim \mathcal{N}(0,1).\label{eq:logvolar1}
\end{equation}
We rely on the prior setup of \cite{kastner2014ancillarity}, and assume a Gaussian prior for the unconditional mean $\mu_\varsigma\sim\mathcal{N}(0,10)$, a transformed Beta prior for the autoregressive parameter $(\phi_\varsigma+1)/2\sim\mathcal{B}(5,1.5)$, and a Gamma prior on the state innovation variances $\sigma_\varsigma^2\sim\mathcal{G}\left(1/2,1/2\right)$. The prior on the initial state is $\varsigma_{h}\sim\mathcal{N}(\mu_\varsigma,\sigma_\varsigma^2/(1-\phi_\varsigma^2))$.\footnote{{We have experimented with heavy tailed error distributions such as $t$-distributed errors \citep[see, e.g.,][in this context]{carriero2022addressing}. There were no noteworthy differences in predictive accuracy relative to our SV implementation with conditionally Gaussian errors.}}

\inlinehead{Tuning Parameters} We also need to choose priors that govern \texttt{xalm}. Here we use independent Gaussian priors: $\theta_1,\theta_2\sim\mathcal{N}(0,0.1^2)$. The final set of parameters that we need to equip with a prior are the hyperparameters that govern the GP kernel (the unconditional variance $\xi$, and the inverse length scale $\lambda$). We use Gamma priors on both:
\begin{equation*}
    \xi\sim\mathcal{G}\left(a_\xi,a_\xi/b_\xi\right), \quad \lambda\sim\mathcal{G}\left(a_\lambda,a_\lambda/b_\lambda\right).
\end{equation*}
Specifically, we set $a_\xi = 0.5$ and $b_\xi = 1$ to establish a rather vague prior on the unconditional variance of the kernel. It is only weakly informative because we normalize the target and predictor variables before estimation, to have unconditional means and variances equal to zero and one, respectively. {We apply this normalization to avoid unintended influences of any priors due to different scalings of the predictors. This aspect is important both in the context of the global-local prior in BLR, but also considering potential sensitivities of the kernel distance metric.} An inverse transformation is then used to recover predictions in the original scale of the data after estimation. 

Prior information on $\lambda$ is imposed to reduce the risk of overfitting by pushing the inverse length scale to smaller values, in line with \citet{hauzenberger2024gaussian} and \citet{clark2024forecasting}. We achieve this by setting $a_\lambda = 0.5$ and $b_\lambda = 0.1 s_y^2$ where $s_y^2$ is the residual variance of running an auxiliary linear AR$(1)$ model for the normalized dependent variable. Given our normalization, $s_y^2\in(0,1]$, this tightens the prior and shifts its expectation to smaller values when the target variable is more persistent (and vice versa).\footnote{In our empirical application, $s_y^2$ usually takes values between $0.7$ and $0.9$ for output growth, and smaller values between $0.3$ and $0.5$ for inflation, depending on the respective training sample. These numbers imply that our proposed prior scaling puts emphasis on smaller inverse length scales, but is not overly tight. Recall also that the MIDAS weights in $\bm{W}$ implicitly interact with $\lambda$ to determine $\bm{\Lambda}$ in Eq. (\ref{eq:kernel}).} We have found this to be a useful semi-automatic scaling choice in this (time series) context.

\subsection{Posterior and predictive inference}

The joint posterior distribution is not of a well-known form, and we therefore use several Metropolis-Hastings steps within a Gibbs sampling algorithm. Our algorithm cycles through the following steps: 
\begin{enumerate}[leftmargin=*]
\item \textbf{Updating the conditional mean.} Given the conditional variances $\bm \Sigma$ and the predictors $\bm X$, we can update the conditional mean for our three different alternatives: 
\begin{enumerate}[leftmargin = *]
\item \textbf{GP}. In this case, we can sample $\bm f \sim \mathcal{N}\left(\overline{\bm{f}},\overline{\bm{V}}\right)$ from a multivariate Gaussian:
\begin{align*}
        \overline{\bm{f}} &= \mathcal{K}_{\bm{\kappa}}(\bm{X},\bm{X})(\mathcal{K}_{\bm{\kappa}}(\bm{X},\bm{X}) + \bm{\Sigma})^{-1}{\bm{y}}\\
        \overline{\bm{V}} &= \mathcal{K}_{\bm{\kappa}}(\bm{X},\bm{X}) - \mathcal{K}_{\bm{\kappa}}(\bm{X},\bm{X})(\mathcal{K}_{\bm{\kappa}}(\bm{X},\bm{X}) + \bm{\Sigma})^{-1}\mathcal{K}_{\bm{\kappa}}(\bm{X},\bm{X}).
    \end{align*}
\item \textbf{BART}. For this option, we update $\bm f$ by following the sampling procedure proposed in \cite{chipman2010bart}; details are provided in Appendix \ref{app:technical}.
\item \textbf{BLR}. We can sample $\bm \beta$ from its conditional posterior, which is multivariate Gaussian and we have $\bm{f} = \bm{X}\bm{\beta}$.
\end{enumerate}
\item \textbf{Updating the conditional variances.} Given a draw for the conditional mean $\bm f$, the error variance or its prediction can be sampled based on $\bm{\epsilon} = (\bm{y} - \bm{f})$ from:
\begin{enumerate}[leftmargin = *]
\item the well-known inverse Gamma posterior distribution in the case of homoskedastic errors, which can be found in any Bayesian textbook, see, e.g., \citet{koop2003bayesian};
\item through sampling the log-volatilities with filtering algorithms for state space models based on state Eq. (\ref{eq:logvolar1}), see e.g., \citet{kastner2014ancillarity}. We rely on the approximation proposed by \citet{omori2007stochastic}.
\end{enumerate}

\item \textbf{Updating tuning/hyperparameters.} The prior variances from the global-local prior setup for the linear regression coefficients can be updated by sampling from inverse Gamma distributions using an auxiliary representation of the HS. All remaining parameters like the hyperparameters of the GP prior or the MIDAS-parameters for \texttt{xalm} are drawn using Metropolis-Hastings updates with a random walk proposal.
\end{enumerate}

In our empirical work, we iterate the sampling algorithm 12,000 times; then we discard the initial 3,000 as burn-in and use each third of the remaining draws for inference. This yields a set of 3,000 draws to be used for computing any object or moment of interest. {Our algorithm exhibits good convergence properties, as measured with standard MCMC diagnostics such as inefficiency factors.} Further details about the sampling algorithm are provided in Appendix \ref{app:technical}.

\section{Investigating model features using synthetic data}\label{sec:synthetic}
\subsection{Data generating processes (DGPs)}\label{sec:dgps}
In this section, we assess the merits of our proposed model using synthetic data in a controlled environment. Our specified DGPs are inspired by recent literature \citep[see, e.g.,][]{ghysels2007midas, andreou2010regression, ghysels2019estimating, mogliani2021bayesian}. However, we also wish to investigate the performance of the proposed MIDAS specifications (along with a large set of competitors) by assuming not only different variable importance profiles of high-frequency covariates (i.e., whether a DGP is relatively sparse or dense), but also different functional forms for the conditional mean. Consequently, to obtain a rich picture through the Monte Carlo exercise, our DGPs vary along these dimensions.

Across all DGPs, we assume $m = T_H/T_L = 3$, capturing the notion of a quarterly/monthly frequency mismatch we also encounter in our empirical application. Similar to \cite{mogliani2021bayesian} and \cite{babii2022machine}, we assume an independent AR(1) process for each high-frequency indicator, $z_{kt} = \rho_z z_{kt-1/3} + \varepsilon_{z,kt}$, where $\varepsilon_{z,kt} \sim \mathcal{N}(0, 1)$, and we specify a moderate degree of serial correlation with $\rho_z = 0.3$. This degree of persistence is reasonable for a typical monthly macroeconomic variable once it is transformed to stationarity (e.g., the monthly growth rate of industrial production or monthly inflation of an aggregate price series).

To specify $\bm W$, we closely follow the MIDAS literature and use three typical/characteristic weighting schemes --- fast-decaying, hump-shaped, and equal weights for $P_H = 12$ high-frequency lags of $z_{kt}$. For $k = 1, \dots, K$, these weighting schemes are defined by an exponential Almon lag (\texttt{xalm}), where we use $\theta_1 = 0$ and $\theta_2 = -1 \times 10^{-1}$ for fast-decaying weights, $\theta_1 = 5 \times 10^{-1}$ and $\theta_2 = -5 \times 10^{-2}$ for hump-shaped weights, and $\theta_1 = \theta_2 = 0$ for equal weights. We map the high-frequency covariates to the lower frequency as specified in Section \ref{sec:econometrics}: $\tilde{\bm {x}}_t = (\bm I_K \otimes \bm W')\tilde{\bm {z}}_t$, with $\tilde{\bm{z}}_t = (\tilde{\bm{z}}_{1t}',\hdots,\tilde{\bm{z}}_{Kt}')'$ and $\tilde{\bm{z}}_{kt} = (z_{kt},z_{kt-1/3},z_{kt-2/3},\hdots,z_{kt-11/3})'$.

To relate our low-frequency endogenous variable $y_t$ to the low-frequency predictors $\tilde{\bm{x}}_t$, we assume the following autoregressive distributed lag (ADL) process:
\begin{equation*}
y_t = \rho_y y_{t-1} + f(\tilde{\bm {x}}_t) + \epsilon_{t}, \quad \epsilon_{t}\sim\mathcal{N}(0,\sigma^2), 
\end{equation*}
with $\epsilon_{t}$ denoting homoskedastic errors with $\sigma^2 = 0.5$. The parameter $\rho_y$ introduces persistence in $y_t$. Similar to our high-frequency indicators, we assume $\rho_y = 0.3$. For $f(\tilde{\bm {x}}_t)$, we vary the functional form from highly nonlinear (\texttt{NL}) to linear (\texttt{L}): 
\begin{itemize}
    \item The nonlinear function is inspired by the \cite{friedman1991multivariate} function and takes the form $f(\tilde{\bm {x}}_t)  = \beta_1 \sin(\pi \times \tilde{x}_{1t} \tilde{x}_{2t}) + 2\beta_1 (\tilde{x}_{3t} - 0.5)^2 +  \beta_2 \tilde{x}_{4t} + \beta_3 \tilde{x}_{5t} + \sum_{j = 6}^{K} 0 \times \tilde{x}_{jt}$, where $\beta_1$ (related solely to the nonlinear parts of the function) is sampled from a standard uniform distribution and $\beta_2$ as well $\beta_3$ are sampled from uniform distribution between $-2$ and $2$.
    \item For the linear case, we assume $f(\tilde{\bm {x}}_t) = \sum_{j =1}^{5} \beta_j \tilde{x}_{jt} + \sum_{j = 6}^{K} 0 \times \tilde{x}_{jt}$, where the five nonzero coefficients are sampled from a Gaussian distribution with mean zero and variance $0.25$.
\end{itemize}
Moreover, note that for both cases (\texttt{NL} and \texttt{L}), we assume that only the first five (compressed) predictors are important in defining movements in $y_t$, while all other covariates are considered to be irrelevant. In this sense, by varying $K$, we also vary the degree of sparsity. We use two different settings in terms of model complexity: one small but relatively dense regression with $K = 10$ --- resulting in a degree of sparsity of $50\%$ (five out of ten covariates are important) --- and one medium-sized but relatively sparse regression with $K = 25$ --- resulting in a degree of sparsity of $80\%$ (five out of $25$ covariates are important). In total, we have twelve DGPs with distinct features.

To evaluate the different model specifications, we simulate $R = 50$ replications for each DGP. We obtain (in-sample) $T_H = 750$ (and thus $T_L = 250$) observations. In addition, we simulate one out-of-sample observation at a lower frequency, used for model evaluation. For each DGP simulation, we predict this out-of-sample observation. As loss measures, we use the continuous ranked probability score \citep[CRPS, see][and Section \ref{sec:appdata}]{gneiting2007strictly}, which we average across the $R$ replications for each DGP.

\subsection{Results for artificial data}\label{sec:simres}
Table \ref{tab:CRPSsynthetic} shows CRPSs for the various MIDAS specifications (the results for mean absolute errors, which only measure point forecast accuracy, are very similar and shown in Appendix \ref{app:empirical}). MIDAS variants vary according to the specification of the conditional mean and weighting scheme. The left-hand panel displays results for our proposed GP-MIDAS variants, while the middle and right-hand panels show results with alternative conditional mean specifications (BART and linear). For the exercise using synthetic data, we consider a linear bridge regression (\texttt{BLR-br}) as a benchmark. 

Overall, Table \ref{tab:CRPSsynthetic} demonstrates that our proposed GP specifications perform well and are competitive across all DGPs. For five out of twelve DGP specifications, a GP variant is the overall winner. Particularly for highly nonlinear DGPs, the GP-MIDAS yields large margins compared to the \texttt{BLR-br} benchmark and the competitors. Interestingly, BART shows a somewhat more consistent performance across weighting schemes and the $\mathbb{L} > 1$ cases are useful when using a tree-based approach.

Taking a closer look, we observe that GP-MIDAS specifications, particularly those equipped with an exponential Almon lag (\texttt{xalm}), show the strongest performance for a nonlinear DGP with fast-decaying or hump-shaped weights. It is also noteworthy that since DGPs are simulated with $\mathbb{L} = 1$, for GP-MIDAS it pays off to compress the raw predictors to a parsimonious specification, while the unrestricted GP-MIDAS or the richer specifications with $\mathbb{L} > 1$ are in a sense overparameterized. By contrast, for \texttt{BLR}, the added flexibility when $\mathbb{L} > 1$ can be marginally beneficial in terms of predictive accuracy. This finding captures the notion that a misspecified linear model can partially approximate nonlinearities through a larger information set.

A GP-MIDAS model with the \texttt{xalm} weighting scheme is capable of recovering the truly underlying weight structure, thus improving upon linear benchmarks. For example, the \texttt{NL-fast} DGPs can be considered as the DGP specification featuring the most nonlinearities, as nonlinearities do not only arise from the conditional mean but also arise from a highly nonlinear weighting scheme implied by the fast-decaying weights. In such a case, a linear bridge regression is extremely misspecified, while a \texttt{GP-xalm} detects those nonlinearities, resulting in large predictive gains. Conversely, when DGPs are linear (which is when the nonparametric versions will be less efficient than BLR by construction), both GP and BART still show very competitive predictive metrics. This implies that our setup achieves sufficient regularization, and alleviates concerns about overfitting.

\begin{table}[t]
\includegraphics[width=\textwidth]{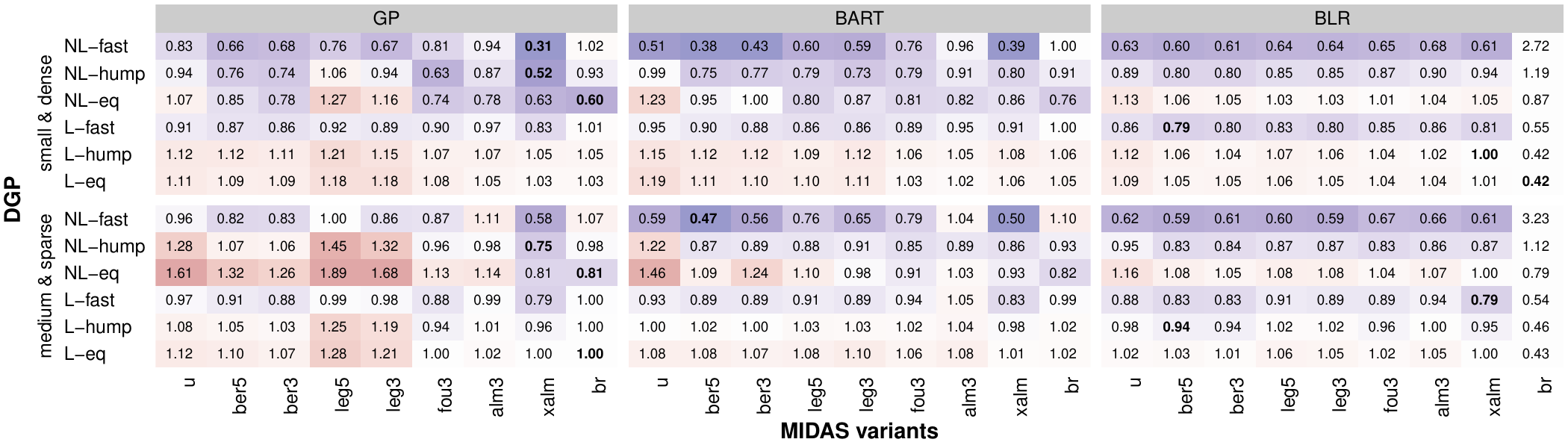}
\caption{Comparison of predictive (nowcast) losses using CRPS against the \texttt{BLR}-\texttt{br} benchmark across different model specifications and data generating processes (DGPs). \textit{Notes}: Each row represents a DGP, columns denote model specifications. DGPs vary the functional form --- \texttt{NL} (nonlinear) vs. \texttt{L} (linear), the weighting scheme --- fast-decaying (\texttt{fast}), hump-shaped (\texttt{hump}), vs. equal (\texttt{eq}) weights, and model complexity --- medium ($K = 25$) and sparse ($80\%$ sparsity) vs. small ($K = 10$) and (relatively) dense ($50\%$ sparsity). The best-performing specification for each DGP is highlighted in bold.}
\label{tab:CRPSsynthetic}
\end{table}

\section{Nowcasting and forecasting output and inflation in the US}\label{sec:application}
\subsection{Data, competing models and predictive comparisons}\label{sec:appdata}
The quarterly target series are sourced from FRED-QD and the monthly predictors are from the FRED-MD database \citep[see][]{mccracken2016fred,mccracken2020fred}. Our target variables are real GDP growth (\texttt{GDPC1}) and inflation in the GDP deflator (\texttt{GDPCTPI}). Both are used in the form of annualized growth rates. Our full sampling period runs from 1963Q1 to 2023Q2, and in our out-of-sample evaluation scheme, the initial estimation (training) sample ends in 1989Q4. For nowcasts and forecasts, the first targeted quarter in our holdout sample is 1990Q1. Charts of the target variables and subsamples we refer to below are provided in Appendix \ref{app:empirical}.\footnote{\texttt{Full} denotes the whole holdout; sample splits are \texttt{Pre Covid} and \texttt{Post Covid}, where the full sample is split at 2019Q4. We further differentiate \texttt{Recession} versus \texttt{Expansion} by considering all dates qualified as falling into a recessionary period by NBER as the former, and the rest of the sample as the latter.}

In the out-of-sample evaluation scheme, we rely on final vintage data, but truncate each of the training samples according to the release calendar as we simulate going forward in time. That is, in our ``pseudo real-time'' evaluation setup, we abstract from data revisions but respect the publication schedule of the variables. We focus on the current quarter nowcast indexed $h \in \{0,1/3,2/3\}$ and the one-quarter-ahead forecast $h \in \{1,4/3,5/3\}$ --- both are updated at the end of each month during a quarter, and $h$ encodes the distance to the target quarter in months. To give an example, a nowcast for 2023Q1 made in 2023M2 (which is referred to as $h = 1/3$), due to publication delay, uses monthly information up to 2023M1 \citep[see also][]{babii2022machine}. A typical monthly predictor would thus take the form $\tilde{\bm{z}}_{kt} = (z_{kt-2/3},z_{kt-1},\hdots,z_{kt-(P_H+1)/3})'$. More generally, for any $h$, we have $\tilde{\bm{z}}_{kt} = (z_{kt-h-1/3},z_{kt-h-2/3},\hdots,z_{kt-h-(P_H)/3})'$.

Potential predictors are defined based on information sets of different sizes: \textit{small} (\texttt{s}, $K=12$), \textit{medium} (\texttt{m}, $K=23$) and \textit{Big Data} (\texttt{b}, $K=116$). These sets are constructed so that the respective larger one nests the smaller ones. A detailed description of which predictors are used in each of them is provided in Appendix \ref{app:empirical}. All these predictor variables are processed using the suggested transformation codes to achieve approximate stationarity. We use $P_L = 4$ lags of the target variable as is common with quarterly data; and, reflecting the month/quarter frequency mismatch, we set $P_H = 12$ based on $P_H = m P_L$ for consistency. In line with \citet{mogliani2021bayesian,mogliani2024bayesian} and \citet{babii2022machine} we choose $\mathbb{L} = 3$ if applicable, and also consider $\mathbb{L} = 5$ for Bernstein and Legendre polynomials.

Besides differently sized information sets, our set of competing models is defined as follows. We vary the respective approach to estimating the conditional means $f(\bullet)$ by using linear (BLR), GP and BART versions. In addition, we consider homoskedastic (\texttt{hom}) and heteroskedastic (\texttt{sv}) implementations. The final variation is due to how we construct the design matrices for distinct MIDAS types. Here, we use both UMIDAS and the restricted approaches we discussed in Section \ref{sec:econometrics_midas}. As a simplistic benchmark, we also estimate univariate $AR(P_L)$ models which are updated each quarter. An overview is provided in Table \ref{tab:competitors}. When we explicitly refer to a model in the text, we structure their IDs as \texttt{mean-variance-midas-size}.

\begin{table}[t]
\caption{Design of the forecast exercise and competing models.}\label{tab:competitors}\vspace*{-0.5em}
\centering\small
\begin{threeparttable}
\begin{tabular*}{\textwidth}{@{\extracolsep{\fill}} lll}
\toprule
\textbf{MIDAS} & \textbf{Mean} & \textbf{Variance}\\
\midrule
Unrestricted, $\mathbb{L} = P_H$ (\texttt{u}) & \multirow{4}{6cm}{Linear (\texttt{BLR}): $f(\bm{x}_{t}) = \bm{x}_{t}'\bm{\beta}$ \\ \texttt{GP}: $f(\bm{x}_{t}) \sim \mathcal{GP}\left(0,\mathcal{K}_{\bm{\kappa}}(\bm{x}_{t},\bm{x}_{t})\right)$ \\ \texttt{BART}: $f(\bm{x}_{t}) \approx \sum_{s=1}^S \ell_{s}(\bm{x}_{t}|\mathcal{T}_s,\bm{\mu}_s)$}
 & \multirow{4}{3cm}{$\sigma_t^2 = \sigma^2$ (\texttt{hom})\\ $\sigma_t^2$ (\texttt{sv})} \\
$\mathbb{L} = 1$ (\texttt{br}, \texttt{xalm}) & &  \\
$\mathbb{L} = 3$ (\texttt{alm}, \texttt{ber}, \texttt{leg}, \texttt{fou}) & & \\
$\mathbb{L} = 5$ (\texttt{ber}, \texttt{leg}) & & \\
\bottomrule
\end{tabular*}
\begin{tablenotes}[para,flushleft]
\scriptsize{\textit{Notes}: All specifications target real GDP growth (\texttt{GDPC1}) or the GDP deflator (\texttt{GDPCTPI}) with small (\texttt{s}, $K=12$), medium (\texttt{m}, $K=23$), big data (\texttt{b}, $K=116$) information sets for nowcasts $h\in\{0,1/3,2/3\}$ and forecasts $h\in\{1,4/3,5/3\}$. Number of predictors: $M = P_L + K (\mathbb{L} + 1)$, which results in approximately $1.4$k predictors in our most flexible specification.}
\end{tablenotes}
\end{threeparttable}
\end{table}

\inlinehead{Loss Functions} We evaluate our predictions using several distinct loss functions that measure different aspects of predictive accuracy. Specifically, we use the quantile score (QS) and the (quantile-weighted) continuous ranked probability score (CRPS), see \citet{giacomini2005evaluation,gneiting2007strictly}. 

The QS for quantile $\tau\in(0,1)$ of the forecast of target variable $y_t$ is defined as $\text{QS}_{\tau,t} = 2\left(y_{t}-\hat{y}_{\tau,t}\right)\left(\tau - \mathbb{I}\left\{y_{t}\leq \hat{y}_{\tau,t}\right\}\right)$, where $\hat{y}_{\tau,t}$ indicates the $\tau$th quantile of the predictive distribution. The indicator function $\mathbb{I}\left\{y_{t}\leq \hat{y}_{\tau,t}\right\}$ has a value of $1$ if the realized value is at or below the quantile forecast, and $0$ otherwise. Note that $\text{QS}_{0.5,t}$ is the absolute error (which we use in the form of the mean absolute error, MAE, as our point forecast loss). Using the QS, we can then define the (quantile-weighted) CRPS following \citet{gneiting2011comparing}:
\begin{equation*}
\text{CRPS-V}_{t} = \int_{0}^{1} \mathfrak{w}_{\text{V},\tau} \text{QS}_{\tau,t} \text{d}\tau, \quad \text{for: V} \in\{\text{\texttt{L},\texttt{R}}\}, 
\end{equation*}
where $\mathfrak{w}_{\text{V},\tau}$ indicate weights that emphasize different parts of the distribution. The default CRPS results when we use equal weights, but we also consider different weighting schemes that target the left tail $\mathfrak{w}_{\text{\texttt{L}},\tau} = (1-\tau)^2$ and the right tail $\mathfrak{w}_{\text{\texttt{R}},\tau} =  \tau^2$. We approximate the integral above with a sum based on a grid of quantiles, $\tau\in\{0.05,0.06,\hdots,0.94,0.95\}$.


\subsection{A bird's-eye view of model performance}
The set of competing models is vast. We therefore slice our empirical results along several dimensions to begin with a concise overview before drilling deeper into details. 

\subsubsection{Which model features are important?}
First, we take a bird's-eye view of which broad model features --- on average --- produce better forecasts. We choose the CRPS averaged for various subsamples over the holdout as our metric of choice for measuring overall forecast performance by target variable. The logarithms of these losses across specifications are then regressed on dummies comprised of the categories determined by the underlying means, variances, sizes of the information set, and MIDAS type (i.e., our broad model features). 

Our ``baseline categories'' reflect the linear homoskedastic bridge model and the small information set --- which are the categories left out by the groups of dummies. The parameters are estimated with OLS and multiplied by $100$. These are the numbers shown in Figure \ref{fig:exploratory}, which can be interpreted as percentage gains/losses due to the extensions indicated in the rows of each panel relative to the losses of the benchmark. {Panel (a) shows the results including all dummies, panel (b) subsets to only GP implementations and thus omits the dummy for the mean component.} The columns refer to the nowcast/forecast horizon $h$.\footnote{When using the mean absolute error (MAE) as explained loss instead, we obtain qualitatively very similar patterns.}

\begin{table}[t]
    \centering
    \begin{subfigure}[t]{\textwidth}
        \caption{\textbf{All specifications}: mean, MIDAS, size, and variance dummies}
        \includegraphics[width=\textwidth]{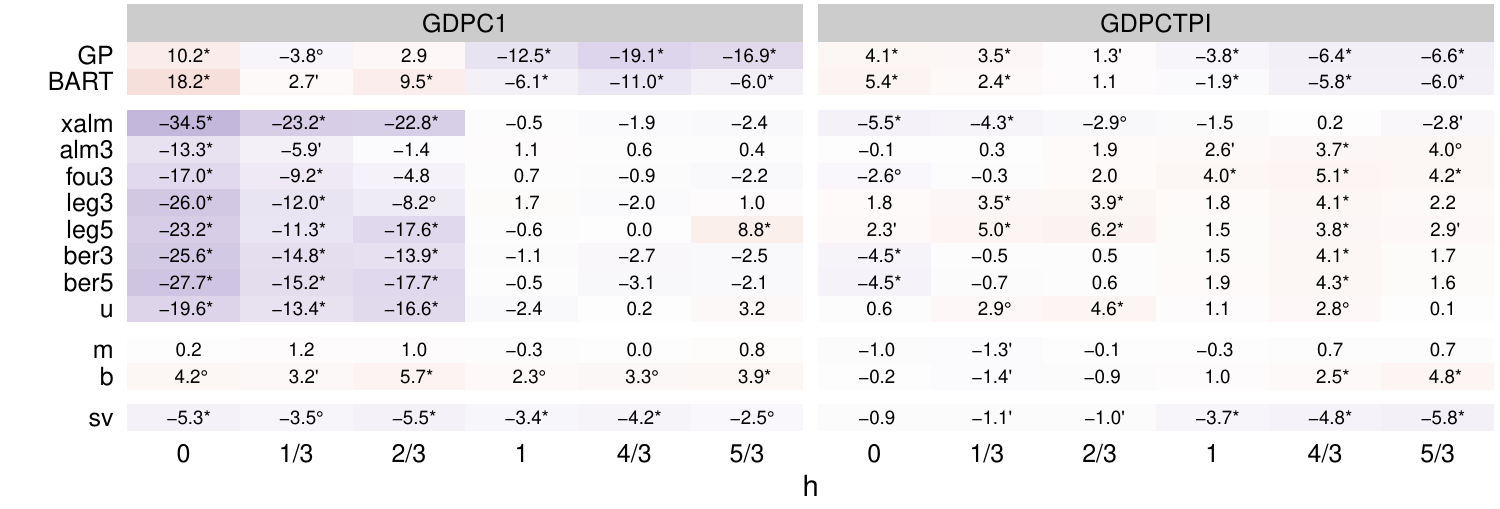}
    \end{subfigure}
    \begin{subfigure}[t]{\textwidth}
        \caption{\textbf{GP specifications}: MIDAS, size, and variance dummies}
        \includegraphics[width=\textwidth]{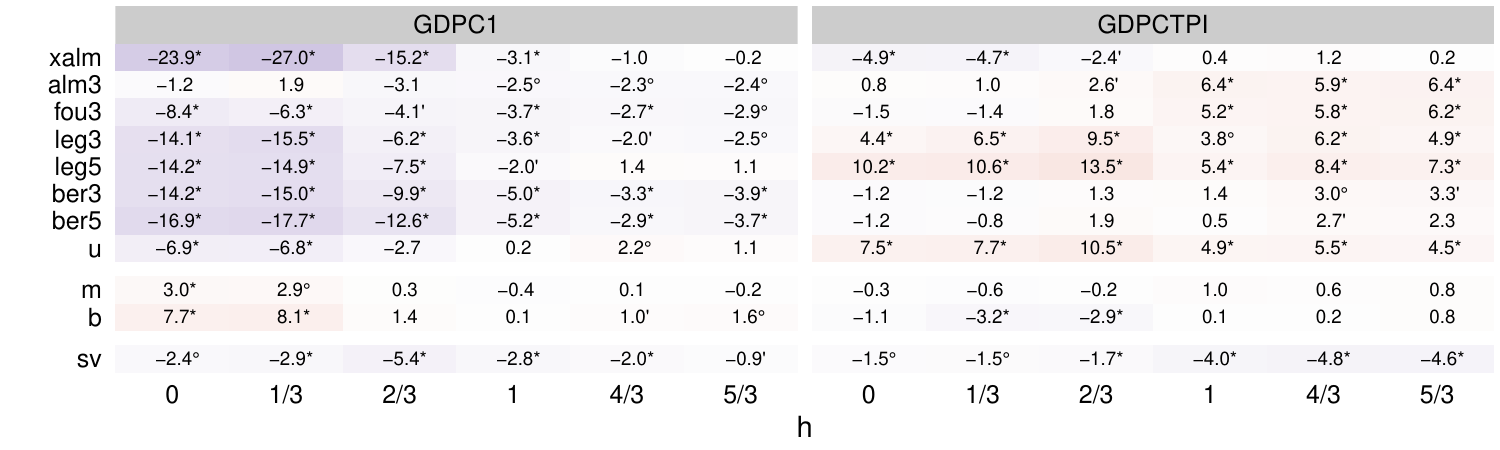}
    \end{subfigure}
    
    \caption{Predictive gains/losses, varying model features by target variable and horizon. \textit{Notes}: Regressions of log-CRPS on dummies. Statistical significance $\{$5, 1, 0.1$\}$\% indicated with $\{',^\circ,^\ast\}$.}
    \label{fig:exploratory}
\end{table}

Starting with panel (a), there is heterogeneity concerning the predicted variable. Addressing choices about the conditional mean first, we find that the nonparametric models pay off particularly as $h$ becomes larger. By contrast, for the nowcasts, we find that GP and BART are often performing worse than when assuming linearity a priori (sometimes even significantly so). It is worth noting, however, that this overview abstracts from interactions of model features. So even when on average losses are larger when moving from BLR to GP or BART, keeping everything else fixed, this is not necessarily the case for distinct model specifications. Indeed, we zoom into model-specific performance below and find that several versions of the GP implementation are the overall best-performing models.

Adjusting the MIDAS weights provides the greatest relative gains, at least for GDPC1. The two-parameter specification \texttt{xalm} seems to succeed in striking a balance between flexibility and simplicity. For GDPCTPI, this aspect is somewhat less important. The corresponding accuracy premium appears consistently across horizons and target variables, although the magnitudes differ somewhat. Another specification detail that consistently improves scores is \texttt{sv}. Usually, these improvements are larger as the forecast horizon increases. The size of the information set does not matter much, although it must be said that using \texttt{b} sometimes significantly hurts predictive accuracy by a few percentage points. These general patterns also apply when subsetting to GP implementations in panel (b), with differentials due to varying these model features being somewhat more muted.

\subsubsection{Which models perform best overall?}
Our second bird's-eye view set of results aims at identifying those specifications that perform best consistently (meaning across all now-/forecast horizons and all subsample splits). To answer the question posed in the title of this subsection, we compute the model confidence set (MCS) of \citet{hansen2011model} for all possible combinations of losses, horizons, and subsamples.

{The output of the MCS procedure is a subset of models which includes the best-performing one with a specified level of confidence. The specifications included in the MCS are statistically indistinguishable at this pre-defined level. Our results are based on the T$_{R}$ statistic defined in \citet{hansen2011model} and we use a $10$ percent level of significance. This procedure provides us with a framework to distinguish models that consistently perform well, as opposed to ones that are strictly dominated by others. Having computed the MCS across all subsets of results, we count how often each model was among the superior set of models, and present inclusion percentages alongside the minimal average loss (across horizons and subsamples) for the best three specifications in Table \ref{tab:mcs}.}

\begin{table}[t]
    \centering
    \begin{subfigure}[t]{0.49\textwidth}
        \caption{\texttt{GDPC1}}
        \includegraphics[width=\textwidth]{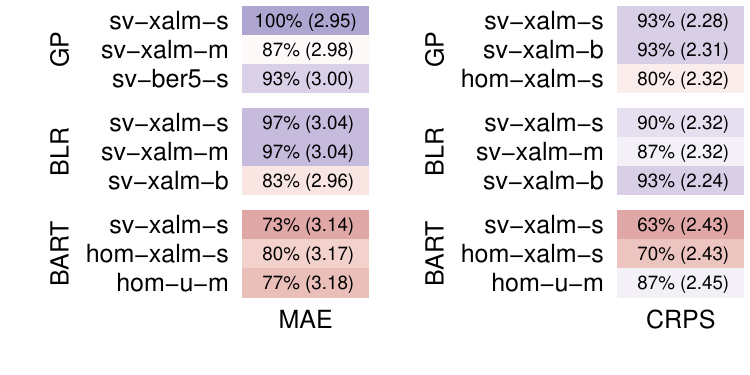}
    \end{subfigure}
    \begin{subfigure}[t]{0.49\textwidth}
        \caption{\texttt{GDPCTPI}}
        \includegraphics[width=\textwidth]{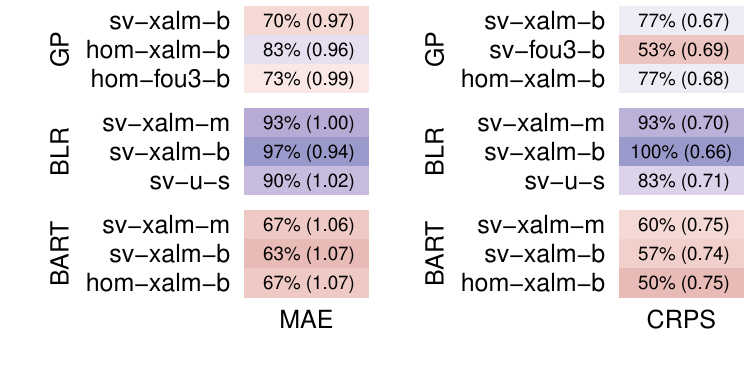}
    \end{subfigure}
    
    \caption{Summary of model confidence sets across horizons and subsamples. \textit{Notes}: Percentages indicate how often the respective specification was among the superior model set, the number in parentheses are averaged losses.}
    \label{tab:mcs}
\end{table}

Again there are differences between target variables, but also, with respect to predictive loss metrics. Consistent with our exploratory regressions, \texttt{xalm} features very prominently, irrespective of how the conditional mean is specified or how large the information set is. Interestingly, short-horizon predictions of inflation as measured by the GDP deflator tend to require larger models (the Big Data information set appears more often) than for real GDP growth, where small and medium models are more common.

One clear lesson of this exercise is that BART is dominated (both in terms of averaged losses and selection frequency) by GP and BLR, and the latter two often exhibit similar performance. For GDPC1, we find \texttt{GP-sv-xalm-s} to perform well very consistently. This model has the lowest average losses for both point forecasts (as measured by the MAE) and the density forecasts captured by CRPSs and is included in the MCS in all cases for the former metric, and 93\% of all cases for the latter. By contrast, for GDPCTPI, while the average loss metrics for the best-performing model specifications are very close in many cases, the MCS favors BLR. As we will see below, this is due to heterogeneities in the performance of GP both across horizons and across subsamples.

\subsubsection{Computation times}
Before proceeding with a more granular discussion of our nowcast and forecast results, it is also worth briefly discussing estimation times. Theoretical considerations about the computational complexity were alluded to in Section \ref{sec:econometrics_mean}; below we provide a summary based on estimation times for our full out-of-sample forecast simulation.\footnote{{This simulation was run in parallel on a high-performance computing cluster with Intel Xeon Platinum 8358 32C 2.6GHz CPUs, a total number of 1,792 cores and 16,896GB of RAM.}} It is worth reiterating that the bottleneck for BLR in its baseline implementation is the number of predictors $M$, while for GP the relevant dimension is the number of observations $T_L$. When we employ a fast-sampling algorithm for BLR (we use this approach when $T_L < M$ because the computational burden for some of our Big Data models is otherwise insurmountable, especially in a recursive prediction exercise), the limiting factor for BLR is also $T_L$. Due to how BART models the conditional mean, neither of the two dimensions is particularly important.

\begin{figure}[!t]
    \centering
    \includegraphics[width = \textwidth]{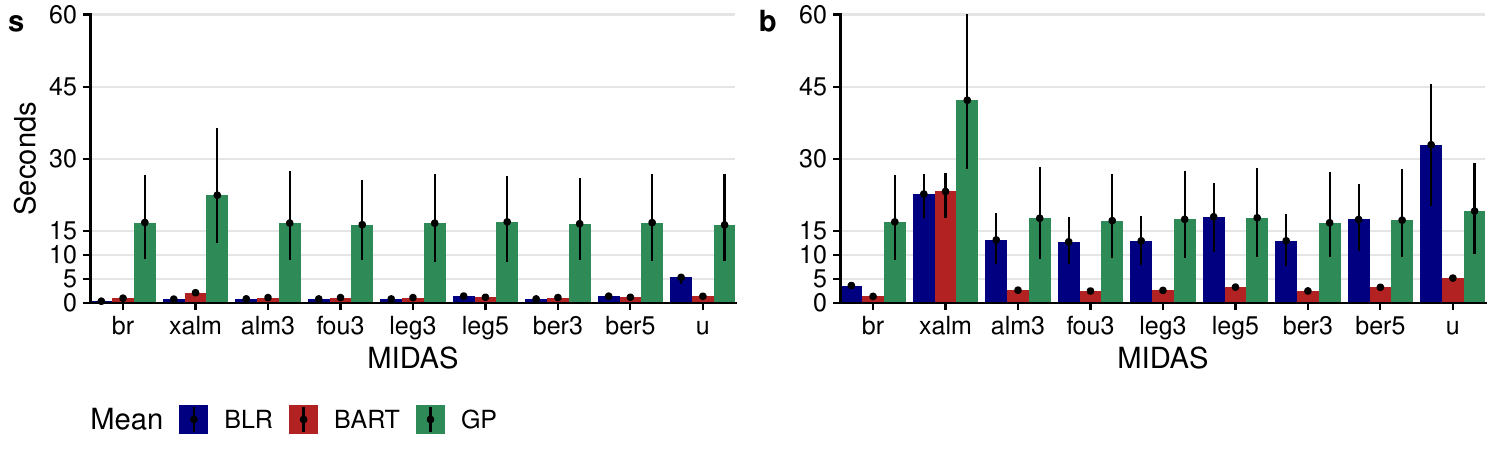}
    \caption{Average estimation times for producing 1,000 draws. Bars indicate the median duration black lines mark $25$th and $75$th percentiles of estimation times across all training samples, small (s) and Big Data (b) information sets.}
    \label{fig:est-times}
\end{figure}

{Figure \ref{fig:est-times} shows the average duration of producing 1,000 draws (in seconds). Since we use an expanding window, the number of observations differs for each training and holdout sample, which is why we indicate the median with colored bars. The black error bars (lines) mark the $25$th and $75$th percentiles of estimation times across all training samples. These results are based on models featuring SV (they are roughly the same for the homoskedastic versions).} 

A few aspects are worth noting. Disregarding \texttt{xalm}, it always takes virtually the same amount of time to estimate the GP specification regarding the size of the information set and polynomial degree. This is due to the number of observations being the constraint, which is reflected in rather wide bars across differently-sized training samples. The same is true when $M > T$ for BLR due to the properties of the fast-sampling algorithm we use in this case. It is worth mentioning that when using a default algorithm for BLR, the scales of this chart would be obscured because BLR turns out to be computationally infeasible as the number of variables becomes huge. At least for the Big Data information set, it roughly takes the same amount of time to estimate GP and BLR, and GP is computationally more efficient when using UMIDAS (i.e., when the number of predictors is huge). For BART, the number of predictors solely affects the number of potential splitting variables, which is why it is on average the fastest of our competitors across model variations. 

{It is also worth noting why the \texttt{xalm} implementation is somewhat slower than the others. This is because it requires several comparatively costly matrix operations in the context of the MH updates. If one were to calibrate the parameters $\theta_1,\theta_2$ beforehand, the computational burden would be roughly the same as in the other cases.} 

\subsection{Model-specific results}
To drill deeper into model-specific predictive accuracy, we select a subset of models that perform best on average as we did for the MCS. We now benchmark these results relative to an AR$(P_L)$-model which is a useful baseline for short-horizon forecasts of output growth and inflation. It reflects the case where no within-quarter information is used. Relative losses are shown as ratios in Table \ref{tab:crps_selected}, and the models are selected based on the sum of the CRPS across all horizons being minimal. We then display the best three GP models and the respective best BLR and BART one, for each target variable. Below we mostly focus on discussing the CRPS, since qualitative differences between point and density forecasts are often muted. Additional metrics for other specifications are collected in Appendix \ref{app:empirical}.

\begin{table}[t]
    \begin{subfigure}{\textwidth}
        \centering\caption{\texttt{GDPC1}}
        \includegraphics[width=0.49\textwidth]{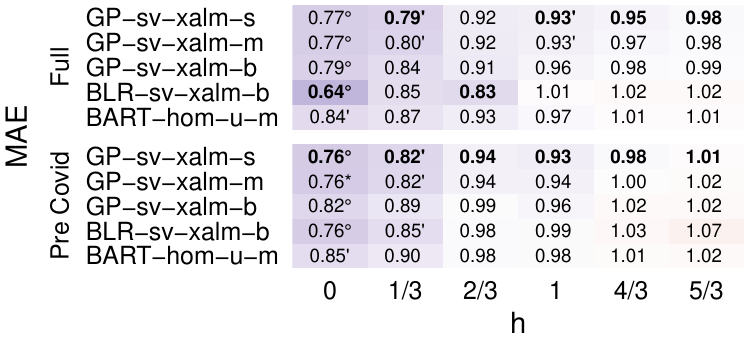}
        \includegraphics[width=0.49\textwidth]{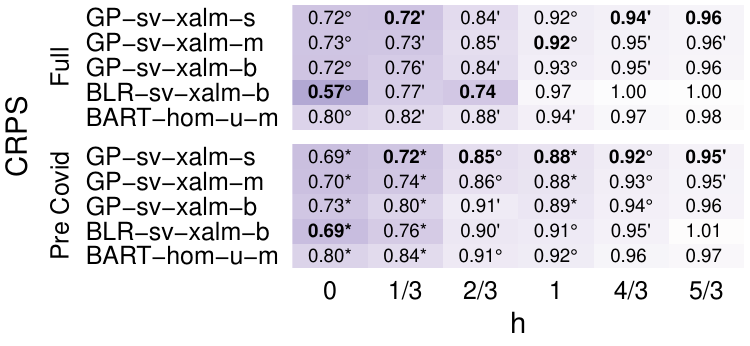}
    \end{subfigure}
    \begin{subfigure}{\textwidth}
        \centering\caption{\texttt{GDPCTPI}}
        \includegraphics[width=0.49\textwidth]{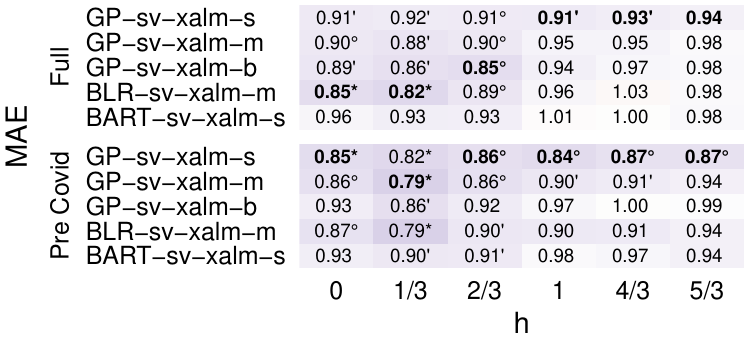}
        \includegraphics[width=0.49\textwidth]{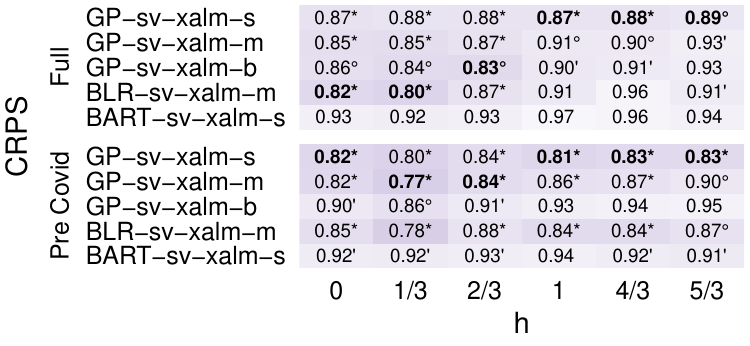}
    \end{subfigure}
    \caption{Predictive losses for selected specifications relative to the benchmark AR($P_L$)-model. \textit{Notes}: Levels of statistical significance $\{$5, 1, 0.1$\}$\% for a DM-test of equal predictive accuracy are indicated with $\{',^\circ,^\ast\}$. Best-performing specification by column in bold.}
    \label{tab:crps_selected}
\end{table}

An obvious starting point when discussing these results is to note that virtually all cells of Table \ref{tab:crps_selected} are colored blue, which is used to indicate that the respective model outperforms the benchmark. In other words, adding within-quarter information for short-horizon predictions --- either when forecasting the next quarter, or when nowcasting the current quarter --- pays off irrespective of whether one addresses any nonlinearities or not. Turning to a discussion of which models end up specifically in this table, we again find MIDAS-type \texttt{xalm} to perform well across all conditional mean specifications, and only in the case of BART, we find a different implementation being included in this set of models. Interestingly, and different from BLR and GP, BART is capable of better extracting information without introducing prior weighting via lag polynomials, though it must be said that this specification is always beaten by another conditional mean implementation. Second, controlling for heteroskedasticity (which we do with \texttt{sv}) pays off. The only exception, again, is BART. This corroborates the findings of \citet{clark2023tail}, who provide a discussion and some empirical evidence of how the regression trees can capture specific forms of heteroskedasticity even when the error terms are homoskedastic.

Focusing on the size of the information set, it is worth mentioning that for the GP it appears that \texttt{s} is sufficient to attain superior relative predictive losses, and the \texttt{b}-sized version only occasionally improves upon the small dataset. By contrast, for BLR, assuming linearity a priori appears to be slightly restrictive, but more information in the form of either the medium or Big Data input appears to help. Another interpretation of this pattern is that the nonparametric features offset omitted variables partly, and the bigger the information set becomes, the less there is a need for modeling nonlinearities explicitly.

A related important observation is that BLR is ranked first (indicated by the bold numbers) in many cases for the nowcasts when considering the full evaluation period. For forecasts, the GP improves upon the linear version for all metrics and both target variables. We conjecture that this is due to the notion that as more information becomes available, the nonlinearities become less emphasized and there is less need for the flexibility in conditional means that our proposed specifications provide. Indeed, this finding is in line with the literature; e.g., \citet{clark2023tail} find that nonparametric models offer larger improvements at longer horizons (where nonlinearities are typically stronger).

When comparing the full holdout sample to the pre-Covid period, we find that the nonparametric models perform better, relatively speaking, when excluding the pandemic observations from the evaluation period. This finding at first seems puzzling, and, in our shorter-horizon predictive context, at odds with the longer-horizon, forecasting-oriented literature. Specifically, earlier related papers have often found more flexible models to offer gains in predictive accuracy in the presence of such outlying observations, due to better robustness properties. Indeed, it is a lack of responsiveness of some of our nonparametric models to huge-variance shocks (reflected timely in the monthly series) that can explain this puzzle. We investigate this claim in the next section, after a more thorough analysis of the resulting predictive distributions.

\subsection{Investigating predictive distributions}
\begin{figure}[t]
    \begin{subfigure}{\textwidth}
        \centering\caption{\texttt{GDPC1}}
        \includegraphics[width=0.49\textwidth]{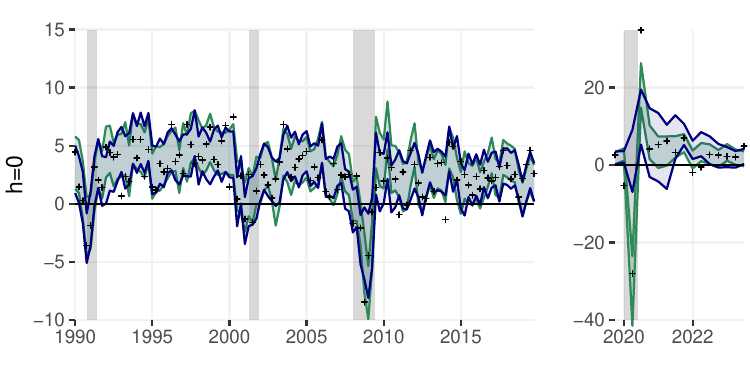}
        \includegraphics[width=0.49\textwidth]{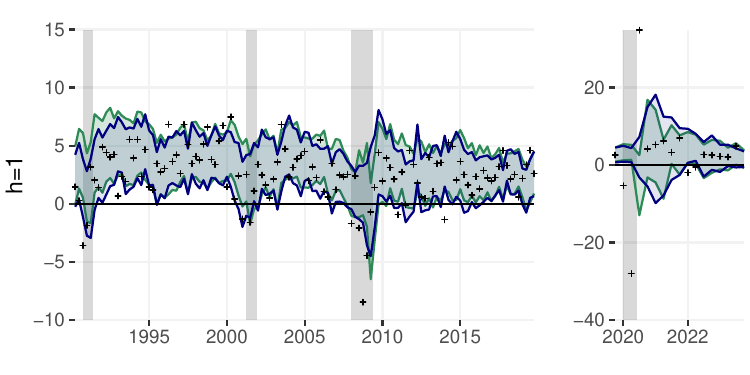}
    \end{subfigure}
    \begin{subfigure}{\textwidth}
        \centering\caption{\texttt{GDPCTPI}}
        \includegraphics[width=0.49\textwidth]{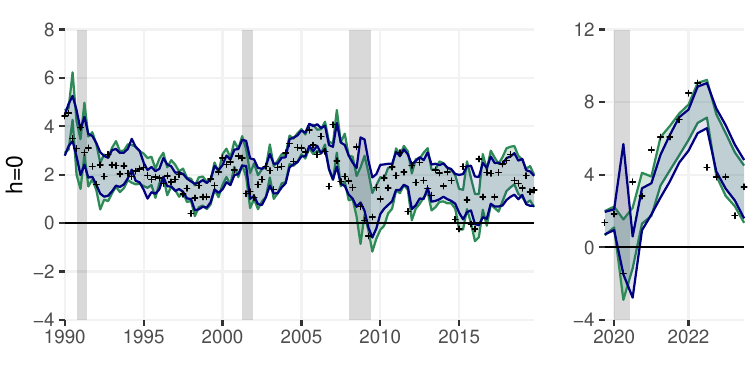}
        \includegraphics[width=0.49\textwidth]{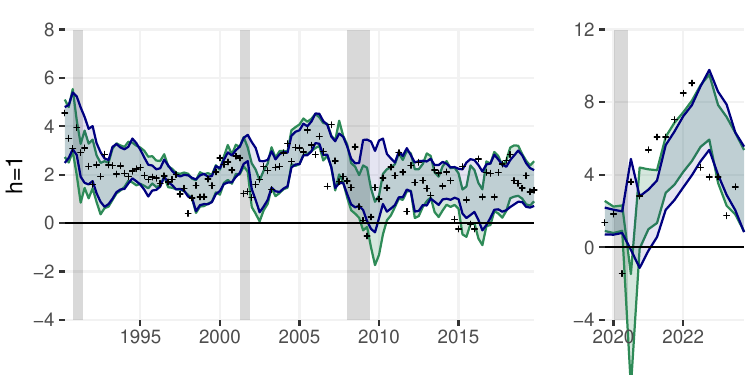}
    \end{subfigure}
    \caption{Predictive distributions over time for the end-of-quarter forecast and nowcast. \textit{Notes}: Black crosses are realized values. The 68 percent predictive credible set is indicated by blue lines and shaded area for \texttt{GP-sv-xalm-s}; green lines and shades mark \texttt{BLR-sv-xalm-m}.}
    \label{fig:pred-dist}
\end{figure}

For the following analysis, we pick the two best-performing models, \texttt{GP-sv-xalm-s} (in blue) and \texttt{BLR-sv-xalm-m} (in green). We refrain from showing BART because it is dominated by the others for the most part. Figure \ref{fig:pred-dist} shows the predictive distributions (68 percent credible set) over time for the end-of-quarter forecast ($h = 1$) and nowcast ($h = 0$) alongside realized values (black crosses).

As one would expect, the credible sets narrow as additional information becomes available to produce the predictions, for both specifications. Deviations between the predicted quantiles, comparing the GP and BLR, are strongest in and around recessionary episodes of the US economy. But we do not detect any noteworthy asymmetries or multimodalities and the like --- features that often arise when related econometric methods are used for forecasting. To some extent, this is by design: we use a \textit{direct} predictive equation and are interested in very short-horizon predictions. So any nonlinearities that might arise from iterative propagation mechanisms and feedback effects \citep[see, e.g.,][]{huber2023nowcasting}, or generally when focusing on higher-order forecasts, do not appear in our out-of-sample exercise. 

Due to this apparent symmetry of the predictive distributions, the tail forecast metrics we consider (CRPS-L for downside risk and CRPS-R for upside risk) are mostly in line with the unweighted CRPS density metric regarding model performance rankings. This is why we report them in Appendix \ref{app:empirical}. One noteworthy pattern is that upside risks in both target variables seem to be captured relatively better by the GP-versions, in line with the findings of \citet{clark2024investigating}. In terms of the magnitudes of improvements over the autoregressive benchmark, the monthly predictors help even more when the interest is on predicting tail risk. Which of the predictors are the most important is what we aim to investigate next.

\inlinehead{Measuring Variable Importance} One shortcoming of machine learning approaches, while usually offering good forecast performance, is that in most cases they lack interpretability. While in a linear regression, the importance of any variable is naturally measured with the size of its regression coefficient, this is not as easily assessed in nonlinear frameworks. To identify which predictors drive the respective predictive distributions we thus resort to auxiliary tools. Specifically, we rely on linearly approximating the relationship between the predictive distribution and the corresponding predictors, inspired by \citet{woody2021model} and applied recently in a similar context to ours by \citet{clark2024forecasting}. 

Let $\hat{y}_{t+h}$ denote the median of the predictive distribution for horizon $h$;\footnote{This procedure may also be used for different quantiles along the predictive distribution. Due to the predictive distributions being mostly symmetric in our empirical exercise, we use the median.} we define the following \textsc{Lasso} problems for each model specification:
\begin{equation}
\hat{\bm{\mathfrak{b}}}_{h} = \min_{\bm{\mathfrak{b}}_{h}} \sum_{t = T_0}^{T} (\hat{y}_{t+h} - \bm{x}_t'\bm{\mathfrak{b}}_{h})^2 + \varrho \sum_{i=1}^M |\mathfrak{b}_{hi}|,\label{eq:lassopred}
\end{equation}
where $T_0$ indexes the first quarter (i.e., 1990Q1) of our holdout sample and $T$ marks the last quarter (2023Q2), $\bm{\mathfrak{b}}_{h} = (\mathfrak{b}_{h1},\hdots,\mathfrak{b}_{hM})'$ are regression coefficients and $\varrho \geq 0$ is a tuning parameter (determined through cross-validation) that governs the weight of the penalty term and degree of sparsity in $\bm{\mathfrak{b}}_{h}$. This procedure recovers a sparse vector $\hat{\bm{\mathfrak{b}}}_{h}$ that measures which predictors in $\bm{x}_t$ shift the location of the predictive distributions (specific to each model variant) throughout the holdout sample.

{We summarize our results for the respective end-of-quarter forecast and nowcast in Table \ref{tab:varimp_both}. Additional results for other horizons are provided in Appendix \ref{app:empirical}. For this exercise, we pick the Big Data variants of BLR and GP, with stochastic volatility and an \texttt{xalm} MIDAS piece. We pick the largest information set for illustration, because these implementations (while not being the overall outright best-performing models at least in the GP case) are competitive, and none of the predictors are excluded a priori. When there are multiple lags of the same variable, we take the sum of the respective coefficients.}

\begin{table}[t]
    \begin{subfigure}[t]{0.49\textwidth}
        \centering\caption{\texttt{GDPC1}}
        \includegraphics[width=\textwidth]{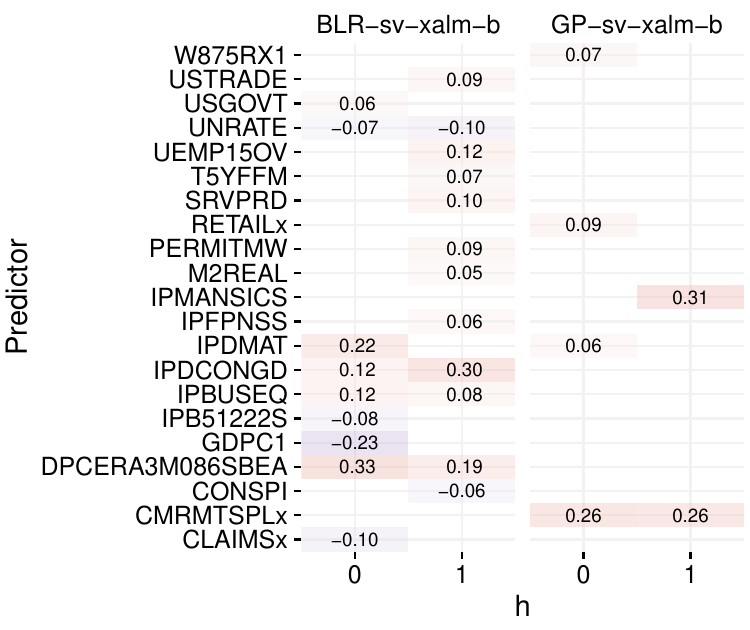}
    \end{subfigure}
    \begin{subfigure}[t]{0.49\textwidth}
        \centering\caption{\texttt{GDPCTPI}}
        \includegraphics[width=\textwidth]{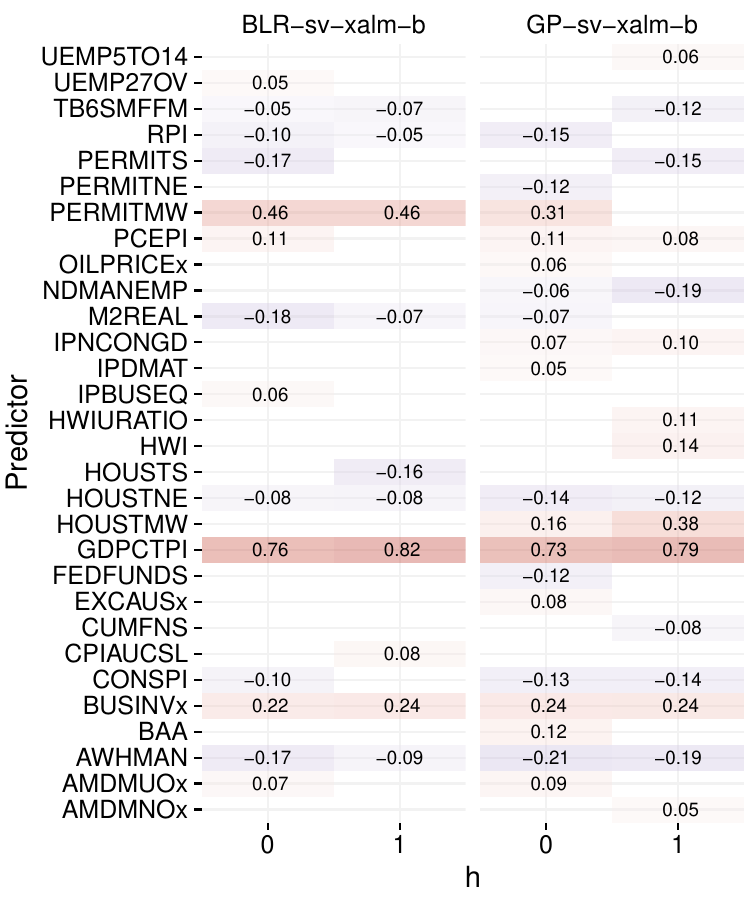}
    \end{subfigure}
    \caption{Most important predictors for both target variables over the holdout for selected model specifications and predictive horizons. \textit{Notes}: Values are standardized coefficients obtained from the \textsc{Lasso} minimization problem given in Eq. (\ref{eq:lassopred}).}
    \label{tab:varimp_both}
\end{table}

A general insight from this approach to measuring variable importance is that models for inflation are denser, in the sense that more predictors shape the forecast distributions, than those for real output growth. This is true for both the linear and nonlinear models. Interestingly, BLR and GP by and large tend to select the same variables, although there are some modest differences in magnitudes of importance. A noteworthy deviation from this pattern occurs for forecasts of output, where GP-MIDAS only has two non-zero predictors while many more are present for BLR. And, while persistence via the autoregressive low-frequency lags is important for inflation, this is not necessarily the case for output.

{Turning to individual predictors, it is noteworthy that components of industrial production (most notably, durable materials and nondurable consumer goods) shift predictions of output growth, as do personal consumption expenditures (\texttt{DPCERA3M086SBEA}) when using BLR. By contrast, GP puts emphasis on manufacturing and trade industry sales (\texttt{CMRMTSPLx}) and industrial production of the manufacturing sectors. For inflation, selected variables are more homogeneous across the two competing specifications, and the GP model relies on a few more predictors. Besides the autoregressive lags of the target variable, labor market variables (e.g., average weekly hours in manufacturing, \texttt{AWHMAN}), business inventories, and variables related to housing (\texttt{HOUST} and \texttt{PERMIT}) seem to matter, as does the real M2 money stock.}

\begin{figure}[t]
    \includegraphics[width=\textwidth]{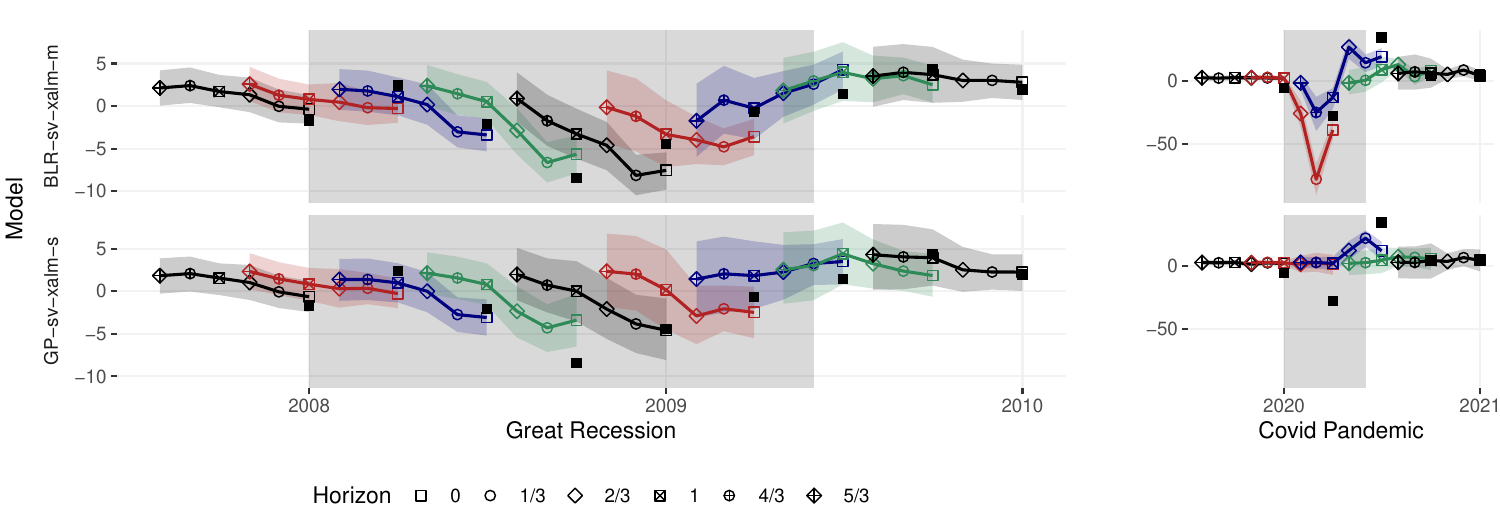}
    \caption{Predictions of real GDP growth during the Great Recession and the Covid Pandemic. \textit{Notes}: Filled black squares mark realized values. Colored lines (predictive median) and shades (68 percent credible set) group forecasts ($h = 5/3, 4/3, 1$) and nowcasts ($h = 2/3, 1/3, 0$) of the same target observation.}
    \label{fig:case-study}
\end{figure}

\inlinehead{A Case Study of Two Recessions} In terms of our earlier note about the role and importance of outlying observations, it is particularly illustrative to focus on two distinct economic episodes: the Great Recession and the Covid pandemic. We do so for real GDP growth in Figure \ref{fig:case-study}, and note that similar arguments can be made in the context of inflation as measured by the GDP deflator. This chart is to be understood as follows. Each of the colored lines and shades refer to the predictive median and 68 percent credible set for a single target observation. Since we update both our forecasts ($h = 5/3, 4/3, 1$) and nowcasts ($h = 2/3, 1/3, 0$) every month, we have six predictions for each quarterly reading of the data (which is marked with a filled black square).

Starting with a review of the Great Recession through the lens of both of these two models, it is noteworthy that during the initial phase of the recession that started in 2008Q1, the prediction paths across horizons are almost identical for both specifications. A key difference then occurs when the trough is reached in 2008Q4: BLR nowcasts this low to persist and overpredicts the severity of the observation in 2009Q1 at $h=0$. By contrast, the GP version is pretty much spot on at the end-of-month horizon for this quarter. These dynamics are due to the nowcast of the nonparametric model responding less strongly to new information in the monthly predictors.

Indeed, an even more prominent example of this phenomenon is visible in the right panel of Figure \ref{fig:case-study}, which shows the quarters most strongly affected by the Covid pandemic. The prediction paths targeting 2020Q2 are especially important for our argument. The difference between the full evaluation and the pre-Covid subsample is essentially due to this period (which is also why the differences are often statistically insignificant): the nonparametric predictions are only very modestly affected by the updated monthly information, while the BLR predictions shift strongly because it linearly extrapolates past patterns. 

For instance, the huge downward spike at $h = 1/3$ (which is the nowcast made in May) reflects the data release for the April reading of the unemployment rate at a historic all-time high and industrial production growth at an all-time low. This lazy reaction of the nonparametric predictions is indeed what provides robustness in a longer-horizon forecasting context, where the goal is to predict future dynamics in periods strictly after a shock materializes. By contrast, when nowcasting, we partly observe the shocks through the monthly variables as they happen. And these more sluggish updates of the GP prediction, as additional information becomes available, can also hurt predictive accuracy in this context.

\section{Conclusions}\label{sec:conclusions}
We develop Bayesian nonparametric methods to be used in combination with MIDAS regressions and discuss their implications. Specifically, we consider GP and BART as flexible alternatives to the penalized linear framework. Our models are also equipped with stochastic volatility, and they are further differentiated with respect to specifics about MIDAS weighting schemes. We show how different MIDAS weights used in GP-MIDAS yield a framework related to compressed GP regression.

After showing their good performance in a set of simulation experiments, the various nonparametric MIDAS specifications are applied to nowcasting and short-term forecasting US real output growth and inflation in the GDP deflator in a large-scale out-of-sample exercise. The set of potential predictors features up to almost $120$ variables for our evaluation period, which starts in the early 1990s. The new models are computationally efficient, and they are competitive while offering gains in predictive accuracy for point, density, and tail forecasts.

{\setstretch{0.9}
\addcontentsline{toc}{section}{References}
\bibliographystyle{custom.bst}
\bibliography{lit}}\normalsize\clearpage\doublespacing

\begin{thebibliography}{52}
\newcommand{\enquote}[1]{``#1''}
\providecommand{\natexlab}[1]{#1}

\bibitem[{Almon(1965)}]{almon1965distributed}
\textsc{Almon S} (1965), \enquote{The distributed lag between capital
  appropriations and expenditures,} \emph{Econometrica} 178--196.

\bibitem[{Andreou \emph{et~al.}(2010)Andreou, Ghysels, and
  Kourtellos}]{andreou2010regression}
\textsc{Andreou E, Ghysels E, and Kourtellos A} (2010), \enquote{Regression
  models with mixed sampling frequencies,} \emph{Journal of Econometrics}
  \textbf{158}(2), 246--261.

\bibitem[{Arro-Cannarsa and Scheufele(2024)}]{arro2024nowcasting}
\textsc{Arro-Cannarsa M, and Scheufele R} (2024), \enquote{Nowcasting GDP: what
  are the gains from machine learning algorithms?} \emph{SNB Working Papers}
  \textbf{6/2024}.

\bibitem[{Babii \emph{et~al.}(2024)Babii, Barbaglia, Ghysels, and
  Striaukas}]{babii2024nowcasting}
\textsc{Babii A, Barbaglia L, Ghysels E, and Striaukas J} (2024),
  \enquote{Nowcasting and Aggregation: Why Small Euro Area Countries Matter,}
  \emph{SSRN} \textbf{4902689}.

\bibitem[{Babii \emph{et~al.}(2022)Babii, Ghysels, and
  Striaukas}]{babii2022machine}
\textsc{Babii A, Ghysels E, and Striaukas J} (2022), \enquote{Machine learning
  time series regressions with an application to nowcasting,} \emph{Journal of
  Business \& Economic Statistics} \textbf{40}(3), 1094--1106.

\bibitem[{Banerjee \emph{et~al.}(2013)Banerjee, Dunson, and
  Tokdar}]{banerjee2013efficient}
\textsc{Banerjee A, Dunson DB, and Tokdar ST} (2013), \enquote{Efficient
  Gaussian process regression for large datasets,} \emph{Biometrika}
  \textbf{100}(1), 75--89.

\bibitem[{Beyhum and Striaukas(2023)}]{beyhum2023factor}
\textsc{Beyhum J, and Striaukas J} (2023), \enquote{Sparse plus dense MIDAS
  regressions and nowcasting during the COVID pandemic,} \emph{arXiv}
  \textbf{2306.13362}.

\bibitem[{Bhattacharya \emph{et~al.}(2016)Bhattacharya, Chakraborty, and
  Mallick}]{bhattacharya2016fast}
\textsc{Bhattacharya A, Chakraborty A, and Mallick BK} (2016), \enquote{Fast
  sampling with Gaussian scale mixture priors in high-dimensional regression,}
  \emph{Biometrika} \textbf{103}(4), 985--991.

\bibitem[{Borup \emph{et~al.}(2023)Borup, Rapach, and
  Sch{\"u}tte}]{borup2023mixed}
\textsc{Borup D, Rapach DE, and Sch{\"u}tte ECM} (2023),
  \enquote{Mixed-frequency machine learning: Nowcasting and backcasting weekly
  initial claims with daily internet search volume data,} \emph{International
  Journal of Forecasting} \textbf{39}(3), 1122--1144.

\bibitem[{Breitung and Roling(2015)}]{breitung2015forecasting}
\textsc{Breitung J, and Roling C} (2015), \enquote{Forecasting inflation rates
  using daily data: A nonparametric MIDAS approach,} \emph{Journal of
  Forecasting} \textbf{34}(7), 588--603.

\bibitem[{Carriero \emph{et~al.}(2015)Carriero, Clark, and
  Marcellino}]{carriero2015realtime}
\textsc{Carriero A, Clark TE, and Marcellino M} (2015), \enquote{Realtime
  nowcasting with a Bayesian mixed frequency model with stochastic volatility,}
  \emph{Journal of the Royal Statistical Society Series A: Statistics in
  Society} \textbf{178}(4), 837--862.

\bibitem[{Carriero \emph{et~al.}(2022{\natexlab{a}})Carriero, Clark, and
  Marcellino}]{carriero2022nowcasting}
---{}---{}--- (2022{\natexlab{a}}), \enquote{Nowcasting tail risk to economic
  activity at a weekly frequency,} \emph{Journal of Applied Econometrics}
  \textbf{37}(5), 843--866.

\bibitem[{Carriero \emph{et~al.}(2022{\natexlab{b}})Carriero, Clark,
  Marcellino, and Mertens}]{carriero2022addressing}
\textsc{Carriero A, Clark TE, Marcellino M, and Mertens E}
  (2022{\natexlab{b}}), \enquote{Addressing COVID-19 outliers in BVARs with
  stochastic volatility,} \emph{Review of Economics and Statistics} 1--38.

\bibitem[{Carvalho \emph{et~al.}(2010)Carvalho, Polson, and
  Scott}]{carvalho2010horseshoe}
\textsc{Carvalho CM, Polson NG, and Scott JG} (2010), \enquote{The horseshoe
  estimator for sparse signals,} \emph{Biometrika} \textbf{97}(2), 465--480.

\bibitem[{Chan \emph{et~al.}(2024)Chan, Poon, and Zhu}]{chan2024tvpmidas}
\textsc{Chan JCC, Poon A, and Zhu D} (2024), \enquote{Time-Varying Parameter
  MIDAS Models: Application to Nowcasting US Real GDP,} \emph{mimeo} .

\bibitem[{Chipman \emph{et~al.}(2010)Chipman, George, and
  McCulloch}]{chipman2010bart}
\textsc{Chipman HA, George EI, and McCulloch RE} (2010), \enquote{BART:
  Bayesian additive regression trees,} \emph{The Annals of Applied Statistics}
  \textbf{4}(1), 266--298.

\bibitem[{Clark \emph{et~al.}(2024{\natexlab{a}})Clark, Huber, Koop, and
  Marcellino}]{clark2024forecasting}
\textsc{Clark TE, Huber F, Koop G, and Marcellino M} (2024{\natexlab{a}}),
  \enquote{Forecasting US inflation using Bayesian nonparametric models,}
  \emph{Annals of Applied Statistics} \textbf{in-press}.

\bibitem[{Clark \emph{et~al.}(2023)Clark, Huber, Koop, Marcellino, and
  Pfarrhofer}]{clark2023tail}
\textsc{Clark TE, Huber F, Koop G, Marcellino M, and Pfarrhofer M} (2023),
  \enquote{Tail forecasting with multivariate bayesian additive regression
  trees,} \emph{International Economic Review} \textbf{64}(3), 979--1022.

\bibitem[{Clark \emph{et~al.}(2024{\natexlab{b}})Clark, Huber, Koop,
  Marcellino, and Pfarrhofer}]{clark2024investigating}
---{}---{}--- (2024{\natexlab{b}}), \enquote{Investigating Growth-at-Risk Using
  a Multicountry Non-parametric Quantile Factor Model,} \emph{Journal of
  Business \& Economic Statistics} \textbf{in-press}.

\bibitem[{Dance and Paige(2022)}]{dance2022fast}
\textsc{Dance H, and Paige B} (2022), \enquote{Fast and Scalable Spike and Slab
  Variable Selection in High-Dimensional Gaussian Processes,} in
  \enquote{International Conference on Artificial Intelligence and Statistics,}
  7976--8002, PMLR.

\bibitem[{Foroni \emph{et~al.}(2015)Foroni, Marcellino, and
  Schumacher}]{foroni2015unrestricted}
\textsc{Foroni C, Marcellino M, and Schumacher C} (2015), \enquote{Unrestricted
  mixed data sampling (MIDAS): MIDAS regressions with unrestricted lag
  polynomials,} \emph{Journal of the Royal Statistical Society: Series A
  (Statistics in Society)} \textbf{178}(1), 57--82.

\bibitem[{Friedman(1991)}]{friedman1991multivariate}
\textsc{Friedman JH} (1991), \enquote{Multivariate adaptive regression
  splines,} \emph{The annals of statistics} \textbf{19}(1), 1--67.

\bibitem[{Ghysels \emph{et~al.}(2024)Ghysels, Marcellino, and
  Valkanov}]{ghysels2024econometric}
\textsc{Ghysels E, Marcellino M, and Valkanov R} (2024), \emph{The econometric
  analysis of mixed frequency data sampling}, Cambridge: Cambridge University
  Press.

\bibitem[{Ghysels and Qian(2019)}]{ghysels2019estimating}
\textsc{Ghysels E, and Qian H} (2019), \enquote{Estimating MIDAS regressions
  via OLS with polynomial parameter profiling,} \emph{Econometrics and
  Statistics} \textbf{9}, 1--16.

\bibitem[{Ghysels \emph{et~al.}(2007)Ghysels, Sinko, and
  Valkanov}]{ghysels2007midas}
\textsc{Ghysels E, Sinko A, and Valkanov R} (2007), \enquote{MIDAS regressions:
  Further results and new directions,} \emph{Econometric Reviews}
  \textbf{26}(1), 53--90.

\bibitem[{Giacomini and Komunjer(2005)}]{giacomini2005evaluation}
\textsc{Giacomini R, and Komunjer I} (2005), \enquote{Evaluation and
  combination of conditional quantile forecasts,} \emph{Journal of Business \&
  Economic Statistics} \textbf{23}(4), 416--431.

\bibitem[{Gneiting and Raftery(2007)}]{gneiting2007strictly}
\textsc{Gneiting T, and Raftery AE} (2007), \enquote{Strictly proper scoring
  rules, prediction, and estimation,} \emph{Journal of the American statistical
  Association} \textbf{102}(477), 359--378.

\bibitem[{Gneiting and Ranjan(2011)}]{gneiting2011comparing}
\textsc{Gneiting T, and Ranjan R} (2011), \enquote{Comparing density forecasts
  using threshold-and quantile-weighted scoring rules,} \emph{Journal of
  Business \& Economic Statistics} \textbf{29}(3), 411--422.

\bibitem[{{Goulet Coulombe} \emph{et~al.}(2022){Goulet Coulombe}, Leroux,
  Stevanovic, and Surprenant}]{goulet2022machine}
\textsc{{Goulet Coulombe} P, Leroux M, Stevanovic D, and Surprenant S} (2022),
  \enquote{How is machine learning useful for macroeconomic forecasting?}
  \emph{Journal of Applied Econometrics} \textbf{37}(5), 920--964.

\bibitem[{Guhaniyogi and Dunson(2016)}]{guhaniyogi2016compressed}
\textsc{Guhaniyogi R, and Dunson DB} (2016), \enquote{Compressed Gaussian
  process for manifold regression,} \emph{Journal of Machine Learning Research}
  \textbf{17}(69), 1--26.

\bibitem[{Hansen \emph{et~al.}(2011)Hansen, Lunde, and Nason}]{hansen2011model}
\textsc{Hansen PR, Lunde A, and Nason JM} (2011), \enquote{The model confidence
  set,} \emph{Econometrica} \textbf{79}(2), 453--497.

\bibitem[{Hastie \emph{et~al.}(2009)Hastie, Tibshirani, and
  Friedman}]{hasti2009elements}
\textsc{Hastie T, Tibshirani R, and Friedman J} (2009), \emph{The Elements of
  Statistical Learning}, NY: Springer New York, 2nd edition.

\bibitem[{Hauzenberger \emph{et~al.}(2024)Hauzenberger, Huber, Marcellino, and
  Petz}]{hauzenberger2024gaussian}
\textsc{Hauzenberger N, Huber F, Marcellino M, and Petz N} (2024),
  \enquote{Gaussian process vector autoregressions and macroeconomic
  uncertainty,} \emph{Journal of Business and Economic Statistics}
  \textbf{in-press}.

\bibitem[{Huber \emph{et~al.}(2023)Huber, Koop, Onorante, Pfarrhofer, and
  Schreiner}]{huber2023nowcasting}
\textsc{Huber F, Koop G, Onorante L, Pfarrhofer M, and Schreiner J} (2023),
  \enquote{Nowcasting in a pandemic using non-parametric mixed frequency VARs,}
  \emph{Journal of Econometrics} \textbf{232}(1), 52--69.

\bibitem[{Kastner and Fr{\"u}hwirth-Schnatter(2014)}]{kastner2014ancillarity}
\textsc{Kastner G, and Fr{\"u}hwirth-Schnatter S} (2014),
  \enquote{Ancillarity-sufficiency interweaving strategy (ASIS) for boosting
  MCMC estimation of stochastic volatility models,} \emph{Computational
  Statistics \& Data Analysis} \textbf{76}, 408--423.

\bibitem[{Kohns and Potjagailo(2023)}]{kohns2023flexible}
\textsc{Kohns D, and Potjagailo G} (2023), \enquote{Flexible Bayesian MIDAS:
  time-variation, group-shrinkage and sparsity,} Staff Working Paper No. 1025,
  Bank of England.

\bibitem[{Koop(2003)}]{koop2003bayesian}
\textsc{Koop G} (2003), \emph{Bayesian econometrics}, Wiley.

\bibitem[{Liu \emph{et~al.}(2020)Liu, Ong, Shen, and Cai}]{liu2020gaussian}
\textsc{Liu H, Ong YS, Shen X, and Cai J} (2020), \enquote{When Gaussian
  process meets big data: A review of scalable GPs,} \emph{IEEE transactions on
  neural networks and learning systems} \textbf{31}(11), 4405--4423.

\bibitem[{Makalic and Schmidt(2015)}]{makalic2015simple}
\textsc{Makalic E, and Schmidt DF} (2015), \enquote{A simple sampler for the
  horseshoe estimator,} \emph{IEEE Signal Processing Letters} \textbf{23}(1),
  179--182.

\bibitem[{Marcellino and Pfarrhofer(2024)}]{marcellino2024chapter}
\textsc{Marcellino M, and Pfarrhofer M} (2024), \enquote{Bayesian Nonparametric
  Methods for Macroeconomic Forecasting,} in \textsc{MP~Clements, and
  AB~Galvao} (eds.) \enquote{Handbook of Macroeconomic Forecasting,} Edward
  Elgar Publishing Ltd.

\bibitem[{McCracken and Ng(2016)}]{mccracken2016fred}
\textsc{McCracken M, and Ng S} (2016), \enquote{FRED-MD: A monthly database for
  macroeconomic research,} \emph{Journal of Business \& Economic Statistics}
  \textbf{34}(4), 574--589.

\bibitem[{McCracken and Ng(2020)}]{mccracken2020fred}
---{}---{}--- (2020), \enquote{FRED-QD: A quarterly database for macroeconomic
  research,} \emph{NBER Working Paper} \textbf{26872}.

\bibitem[{Mogliani and Simoni(2021)}]{mogliani2021bayesian}
\textsc{Mogliani M, and Simoni A} (2021), \enquote{Bayesian MIDAS penalized
  regressions: estimation, selection, and prediction,} \emph{Journal of
  Econometrics} \textbf{222}(1), 833--860.

\bibitem[{Mogliani and Simoni(2024)}]{mogliani2024bayesian}
---{}---{}--- (2024), \enquote{Bayesian Bi-level Sparse Group Regressions for
  Macroeconomic Forecasting,} \emph{arXiv} \textbf{2404.02671}.

\bibitem[{Omori \emph{et~al.}(2007)Omori, Chib, Shephard, and
  Nakajima}]{omori2007stochastic}
\textsc{Omori Y, Chib S, Shephard N, and Nakajima J} (2007),
  \enquote{Stochastic volatility with leverage: Fast and efficient likelihood
  inference,} \emph{Journal of Econometrics} \textbf{140}(2), 425--449.

\bibitem[{Paananen \emph{et~al.}(2019)Paananen, Piironen, Andersen, and
  Vehtari}]{paananen2019variable}
\textsc{Paananen T, Piironen J, Andersen MR, and Vehtari A} (2019),
  \enquote{Variable selection for Gaussian processes via sensitivity analysis
  of the posterior predictive distribution,} in \enquote{The 22nd international
  conference on artificial intelligence and statistics,} 1743--1752, PMLR.

\bibitem[{Pettenuzzo \emph{et~al.}(2016)Pettenuzzo, Timmermann, and
  Valkanov}]{pettenuzzo2016midas}
\textsc{Pettenuzzo D, Timmermann A, and Valkanov R} (2016), \enquote{A MIDAS
  approach to modeling first and second moment dynamics,} \emph{Journal of
  Econometrics} \textbf{193}(2), 315--334.

\bibitem[{Rodriguez and Puggioni(2010)}]{rodriguez2010mixed}
\textsc{Rodriguez A, and Puggioni G} (2010), \enquote{Mixed frequency models:
  Bayesian approaches to estimation and prediction,} \emph{International
  Journal of Forecasting} \textbf{26}(2), 293--311.

\bibitem[{Schnorrenberger \emph{et~al.}(2024)Schnorrenberger, Schmidt, and
  Guilherme}]{schnorrenberger2024harnessing}
\textsc{Schnorrenberger R, Schmidt A, and Guilherme V} (2024),
  \enquote{Harnessing Machine Learning for Real-Time Inflation Nowcasting,}
  \emph{De Nederlandsche Bank WP} \textbf{806}.

\bibitem[{Snelson and Ghahramani(2012)}]{snelson2012variable}
\textsc{Snelson E, and Ghahramani Z} (2012), \enquote{Variable noise and
  dimensionality reduction for sparse Gaussian processes,} \emph{arXiv preprint
  arXiv:1206.6873} .

\bibitem[{Williams and Rasmussen(2006)}]{williams2006gaussian}
\textsc{Williams CK, and Rasmussen CE} (2006), \emph{Gaussian processes for
  machine learning}, volume~2, Cambridge, MA, USA: MIT Press.

\bibitem[{Woody \emph{et~al.}(2021)Woody, Carvalho, and
  Murray}]{woody2021model}
\textsc{Woody S, Carvalho CM, and Murray JS} (2021), \enquote{Model
  interpretation through lower-dimensional posterior summarization,}
  \emph{Journal of Computational and Graphical Statistics} \textbf{30}(1),
  144--161.

\end{thebibliography}

\begin{appendices}\crefalias{section}{appsec}
\begin{center}
{\LARGE\sffamily\textbf{Online Appendix:\\\titletext}}
\end{center}

\setcounter{page}{1}
\setcounter{section}{0}
\setcounter{equation}{0}
\setcounter{footnote}{0}

\renewcommand\thesection{\Alph{section}}
\renewcommand\theequation{\Alph{section}.\arabic{equation}}
\renewcommand\thefigure{\Alph{section}.\arabic{figure}}
\renewcommand\theequation{\Alph{section}.\arabic{equation}}

\section{Technical Appendix}\label{app:technical}
\subsection{Details about MIDAS basis functions}
The $\mathbb{L} = 1$ cases (exponential Almon lags and the bridge model) were described in the main text, alongside non-orthogonalized power polynomials. Other options include Legendre and Bernstein polynomials, as well as Fourier basis functions. The corresponding basis functions are stated below.

\inlinehead{Legendre Polynomials} These are a special case of Jacobi polynomials and were used in \citet{babii2022machine}. We set the unshifted version of the basis functions as:
\begin{equation*}
    \tilde{\varphi}_{l+1}(p) = \frac{2l + 1}{l+1} \cdot p \cdot \tilde{\varphi}_l(p) - \frac{l}{l+1} \cdot \tilde{\varphi}_{l-1}(p),
\end{equation*}
where $\tilde{\varphi}_{0}(p) = 1$ and $\tilde{\varphi}_{1}(p) = p$ and all higher-order degrees are obtained from the recurrence relations presented above. We apply the transformation $\varphi_l(p) = \tilde{\varphi}_l(2p - 1)$ which implies that the polynomials $\varphi_l(p)$ are orthogonal on the interval $[0,1]$.

\inlinehead{Bernstein Polynomials} Inspired by \citet{mogliani2024bayesian}, we use Bernstein basis polynomials to define the underlying basis functions for a polynomial of order $\mathbb{L}$. They are defined as:
\begin{equation*}
    \varphi_{l}(p) = \binom{\mathbb{L}}{l} p^l (1 - p)^{\mathbb{L}-l}.
\end{equation*}

\inlinehead{Fourier Basis} For this option, we set the $\varphi_0(p) = 1$ and then proceed as:
\begin{equation*}
    \varphi_{l}(p) = 
    \begin{cases}
        \cos(l\varpi p) & \text{if } l \text{ is odd},\\
        \sin(l\varpi p) & \text{if } l \text{ is even},
    \end{cases}
\end{equation*}
where \citet{chan2024tvpmidas} suggest setting $\varpi = 2\pi / (\mathbb{L}m)$ due to a favorable alternative representation that arises from this choice in the linear case.

\subsection{Prior setup and posteriors for BLR}
Following \citet{carvalho2010horseshoe}, we establish the HS prior by assuming a Gaussian scale mixture on $\bm{\beta}|\tau_{\beta},\{\lambda_{\beta m}\}_{m=1}^M \sim \mathcal{N}(\bm{0}_M,\underline{\bm{V}}_\beta)$ with $\underline{\bm{V}}_\beta = \tau_{\beta}^2 \cdot \text{diag}(\lambda_{\beta1}^2,\hdots,\lambda_{\beta M}^2)$. The $\tau_{\beta}\sim\mathcal{C}^{+}(0,1)$ and $\lambda_{\beta m} \sim \mathcal{C}^{+}(0,1)$, for $m = 1,\hdots,M,$ are global and local shrinkage parameters and $\mathcal{C}^{+}$ is the half-Cauchy distribution. Another representation allows for more efficient sampling \citep[see][]{makalic2015simple}:
\begin{align*}
    \tau_{\beta}^2|\tilde{\tau}_{\beta}\sim\mathcal{G}^{-1}(1/2,1/\tilde{\tau}_{\beta}), \quad \lambda_{\beta m}^2|\tilde{\lambda}_{\beta m}\sim\mathcal{G}^{-1}(1/2,1/\tilde{\lambda}_{\beta m}), \quad \tilde{\tau}_{\beta},\{\tilde{\lambda}_{\beta m}\}_{m=1}^M \sim \mathcal{G}^{-1}(1/2,1),
\end{align*}
which yields the following posterior distributions:
\begin{align*}
    \tau_{\beta}^2|\bullet&\sim\mathcal{G}^{-1}\left(\frac{M+1}{2}, \frac{1}{\tilde{\tau}_{\beta}} + \sum_{m=1}^M \frac{\beta_m^2}{\lambda_{\beta m}^2}\right), \quad \lambda_{\beta m}^2|\bullet\sim\mathcal{G}^{-1}\left(1,\frac{1}{\tilde{\lambda}_{\beta m}} + \frac{\beta_m^2}{2\tau_{\beta}^2}\right),\\
    \tilde{\tau}_{\beta}|\bullet&\sim\mathcal{G}^{-1}\left(1,1 + \tau_{\beta}^{-2}\right), \quad \tilde{\lambda}_{\beta m}|\bullet\sim\mathcal{G}^{-1}\left(1, 1 + \lambda_{\beta m}^{-2}\right).
\end{align*}
Let $\underline{\bm{y}} = \bm{\Sigma}^{-1/2}\bm{y}$ and $\underline{\bm{X}} = \bm{\Sigma}^{-1/2}\bm{X}$. The posterior distribution of the regression coefficients is:
\begin{align}
    \bm{\beta}|\bullet\sim\mathcal{N}\left(\overline{\bm{\beta}}, \overline{\bm{V}}_\beta\right), \quad \overline{\bm{V}}_\beta = \left(\underline{\bm{V}}_\beta^{-1} + \underline{\bm{X}}'\underline{\bm{X}}\right)^{-1}, \quad \overline{\bm{\beta}} = \overline{\bm{V}}_\beta \underline{\bm{X}}'\underline{\bm{y}}.\label{eq:BLRpost}
\end{align}
In cases when $T_L < M$ it is computationally inefficient to rely on standard procedures for generating random draws from this distribution. Thus, we follow \citet{bhattacharya2016fast} and use the following fast-sampling algorithm: sample $\bm{u}\sim\mathcal{N}(\bm{0}_M,\underline{\bm{V}}_\beta)$ and $\bm{d}\sim\mathcal{N}(\bm{0}_{T_L},\bm{I}_{T_L})$ and set $\bm{v} = \underline{\bm{X}}\bm{u} + \bm{d}$. Subsequently we compute $\bm{w} = (\underline{\bm{X}}\underline{\bm{V}}_\beta\underline{\bm{X}}')^{-1}(\underline{\bm{y}} - \bm{v})$ and set $\hat{\bm{\beta}} = \bm{u} + \underline{\bm{V}}_\beta\underline{\bm{X}}'\bm{w}$. It can be shown that $\hat{\bm{\beta}}$ is a draw from the distribution given in Eq. (\ref{eq:BLRpost}). The computational gains arise from trading the inversion of the $M\times M$ posterior precision for inverting a $T_L\times T_L$-dimensional matrix.

\subsection{Prior setup and posteriors for BART}
Our prior setup for BART closely follows \citet{chipman2010bart}. We set the number of trees $S = 250$, a common choice that works well for many datasets, and rely on regularization. The corresponding priors that achieve such regularization govern a tree-generating stochastic process and comprise three main layers of shrinkage. 

We define the probability of a node at depth $d\in\mathbb{Z}^{+}$ being non-terminal as $a(1 + d)^{-b}$ with $a\in(0,1)$ and $b\in\mathbb{R}^{+}$. Specifically, we again rely on default choices and set $a=0.95$ and $b = 2$. Next, we assume a uniform prior over splitting rules, that is, all predictors are equally likely to determine partitions that yield the tree structure. The threshold values of splitting rules are again assigned a uniform prior over the range of the respective predictor. These assumptions define the prior on the tree structures $p(\mathcal{T}_s)$. 

On the corresponding terminal node parameters $\bm{\mu}_s = (\mu_{s1},\hdots,\mu_{s\#\text{TN}_s})'$ where $\#\text{TN}_s$ is the number of terminal nodes of tree $s$ we assume Gaussian priors. $R_{y} = \max(\bm{y}) - \min(\bm{y})$ denotes the range of $\bm{y}$; we set $\mu_{sj} \sim \mathcal{N}(0,V_\mu)$ for $j = 1,\hdots,\#\text{TN}_s,$ where $V_\mu = R_y^2(4\gamma^2S)^{-1}$. Following precedent and setting $\gamma=2$ implies that about $95$ percent of the prior probability is associated with values inside $R_{y}$.
 
Each tree is sampled conditional on the other $S-1$ trees, which we indicate with $\mathcal{T}_{-s}$, and marginal of the terminal node parameters. That is, we may use a Metropolis-Hastings algorithm to update the trees using the conditional distributions: 
\begin{equation*}
    p(\mathcal{T}_s|\bm{y},\mathcal{T}_{-s},\bullet) \propto p(\mathcal{T}_s) \int p(\bm{y}|\bm{\mu}_s,\mathcal{T}_s,\mathcal{T}_{-s},\bullet)p(\bm{\mu}_s|\mathcal{T}_s) d\bm{\mu}_s,
\end{equation*}
and a transition distribution (where \texttt{new} indicates a proposed tree and \texttt{old} refers to the previously accepted one), $q\left(\mathcal{T}_s^{\text{\texttt{new}}}|\mathcal{T}_s^{\text{\texttt{old}}}\right)$. This transition distribution is defined by four distinct probabilities: growing a terminal node, $\Pr(\text{\texttt{grow}}) = 0.25$; merging terminal nodes, $\Pr(\text{\texttt{merge}}) = 0.25$; change an interior splitting rule, $\Pr(\text{\texttt{change}}) = 0.4$; swap parent and child node, $\Pr(\text{\texttt{swap}}) = 0.1$. The acceptance probability of the new tree is:
\begin{equation*}
    \min\left(\frac{p(\mathcal{T}_s^{\text{\texttt{new}}}|\bm{y},\mathcal{T}_{-s},\bullet)q\left(\mathcal{T}_s^{\text{\texttt{old}}}|\mathcal{T}_s^{\text{\texttt{new}}}\right)}{p(\mathcal{T}_s^{\text{\texttt{old}}}|\bm{y},\mathcal{T}_{-s},\bullet)q\left(\mathcal{T}_s^{\text{\texttt{new}}}|\mathcal{T}_s^{\text{\texttt{old}}}\right)},1\right).
\end{equation*}
Conditional on the trees, the terminal node parameters follow well-known Gaussian posteriors similar to Eq. (\ref{eq:BLRpost}).

\newpage
\section{Empirical Appendix}\label{app:empirical}
\subsection{Data}
\begin{figure}[ht]
    \centering
    \begin{subfigure}[t]{\textwidth}
        \caption{\texttt{GDPC1}}
        \includegraphics[width=\textwidth]{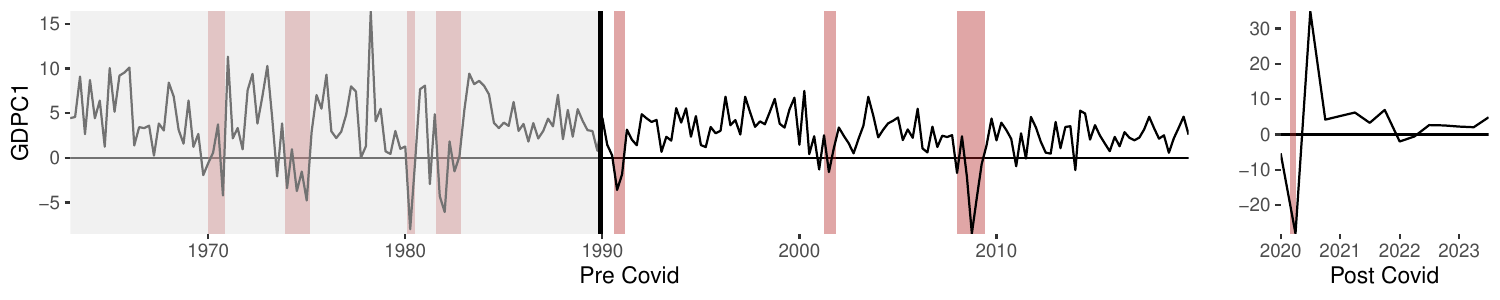}
    \end{subfigure}
    \begin{subfigure}[t]{\textwidth}
        \caption{\texttt{GDPCTPI}}
        \includegraphics[width=\textwidth]{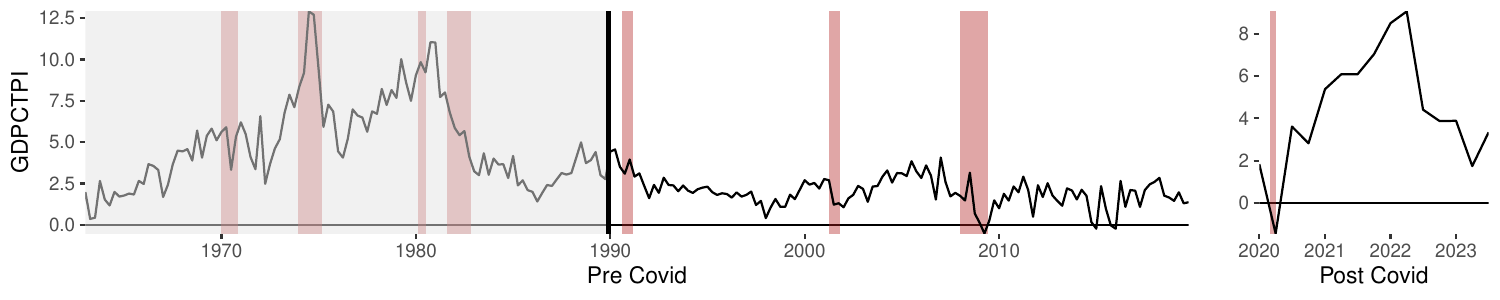}
    \end{subfigure}
    
    \caption{Target variables and visualization of our forecast exercise. \textit{Notes}: The grey shaded area and thick black vertical line delineate our initial training sample. Red shaded areas are NBER recessions.}
    \label{fig:data}
\end{figure}

\begin{table}[ht]
\caption{Data description.}\label{tab:app_data}\vspace*{-0.5em}
\centering\small
\begin{threeparttable}
\begin{tabular}{@{}p{\textwidth}@{}}
\resizebox{\columnwidth}{!}{%
\begin{tabular}{llccc|llccc}
\toprule

\textbf{Code} & \textbf{Description} & \textbf{Trans.} &  &  & \textbf{Code} & \textbf{Description} & \textbf{Trans.} &  &\\
\midrule
\multicolumn{10}{l}{\textit{Quarterly target variables}}\\
\midrule
GDPC1 & Real GDP, 3 Decimal (Billions of Chained 2012 \$) & 8 & & &GDPCTPI & GDP: Chain-type Price Index (Index 2009=100) & 8 &  &  \\
  \midrule
  \multicolumn{10}{l}{\textit{Monthly predictors}}\\
  \midrule
  RPI & Real Personal Income &   5 &  &  & AMDMUOx & Unfilled Orders for Durable Goods &   5 &  &  \\ 
  W875RX1 & Real personal income ex transfer receipts &   5 &  &  & BUSINVx & Total Business Inventories &   5 &  &  \\ 
  DPCERA3M086SBEA & Real personal consumption expenditures &   5 & x & x & ISRATIOx & Total Business:  Inventories to Sales Ratio &   2 &  &  \\ 
  CMRMTSPLx & Real Manu.  and Trade Industries Sales &   5 &  &  & M2SL & M2 Money Stock &   6 &  & x \\ 
  RETAILx & Retail and Food Services Sales &   5 &  & x & M2REAL & Real M2 Money Stock &   5 &  &  \\ 
  INDPRO & IP Index &   5 & x & x & BUSLOANS & Commercial and Industrial Loans &   6 &  &  \\ 
  IPFPNSS & IP: Final Products and Nonindustrial Supplies &   5 &  &  & REALLN & Real Estate Loans at All Commercial Banks &   6 &  &  \\ 
  IPFINAL & IP: Final Products (Market Group) &   5 &  &  & NONREVSL & Total Nonrevolving Credit &   6 &  &  \\ 
  IPCONGD & IP: Consumer Goods &   5 &  &  & CONSPI & Nonrevolving consumer credit to Personal Income &   2 &  &  \\ 
  IPDCONGD & IP: Durable Consumer Goods &   5 &  &  & S\&P 500 & S\&P's Common Stock Price Index: Composite &   5 & x & x \\ 
  IPNCONGD & IP: Nondurable Consumer Goods &   5 &  &  & S\&P: indust & S\&P's Common Stock Price Index: Industrials &   5 &  &  \\ 
  IPBUSEQ & IP: Business Equipment &   5 &  &  & FEDFUNDS & Effective Federal Funds Rate &   2 & x & x \\ 
  IPMAT & IP: Materials &   5 &  &  & TB3MS & 3-Month Treasury Bill: &   2 &  &  \\ 
  IPDMAT & IP: Durable Materials &   5 &  &  & TB6MS & 6-Month Treasury Bill: &   2 &  &  \\ 
  IPNMAT & IP: Nondurable Materials &   5 &  &  & GS1 & 1-Year Treasury Rate &   2 &  &  \\ 
  IPMANSICS & IP: Manufacturing (SIC) &   5 &  &  & GS5 & 5-Year Treasury Rate &   2 &  &  \\ 
  IPB51222s & IP: Residential Utilities &   5 &  &  & GS10 & 10-Year Treasury Rate &   2 & x & x \\ 
  IPFUELS & IP: Fuels &   5 &  &  & AAA & Moody's Seasoned Aaa Corporate Bond Yield &   2 &  &  \\ 
  CUMFNS & Capacity Utilization:  Manufacturing &   2 &  & x & BAA & Moody's Seasoned Baa Corporate Bond Yield &   2 &  &  \\ 
  HWI & Help-Wanted Index for United States &   2 &  &  & COMPAPFFx & 3-Month Commercial Paper Minus FEDFUNDS &   1 &  &  \\ 
  HWIURATIO & Ratio of Help Wanted/No.  Unemployed &   2 &  &  & TB3SMFFM & 3-Month Treasury C Minus FEDFUNDS &   1 &  &  \\ 
  CLF16OV & Civilian Labor Force &   5 &  &  & TB6SMFFM & 6-Month Treasury C Minus FEDFUNDS &   1 &  &  \\ 
  CE16OV & Civilian Employment &   5 &  & x & T1YFFM & 1-Year Treasury C Minus FEDFUNDS &   1 &  &  \\ 
  UNRATE & Civilian Unemployment Rate &   2 & x & x & T5YFFM & 5-Year Treasury C Minus FEDFUNDS &   1 &  &  \\ 
  UEMPMEAN & Average Duration of Unemployment (Weeks) &   2 &  &  & T10YFFM & 10-Year Treasury C Minus FEDFUNDS &   1 &  & x \\ 
  UEMPLT5 & Civilians Unemployed - Less Than 5 Weeks &   5 &  &  & AAAFFM & Moody's Aaa Corporate Bond Minus FEDFUNDS &   1 &  & x \\ 
  UEMP5TO14 & Civilians Unemployed for 5-14 Weeks &   5 &  &  & BAAFFM & Moody's Baa Corporate Bond Minus FEDFUNDS &   1 & x & x \\ 
  UEMP15OV & Civilians Unemployed - 15 Weeks \& Over &   5 &  &  & EXSZUSx & Switzerland / U.S. Foreign Exchange Rate &   5 &  &  \\ 
  UEMP15T26 & Civilians Unemployed for 15-26 Weeks &   5 &  &  & EXJPUSx & Japan / U.S. Foreign Exchange Rate &   5 &  &  \\ 
  UEMP27OV & Civilians Unemployed for 27 Weeks and Over &   5 &  &  & EXUSUKx & U.S. / U.K. Foreign Exchange Rate &   5 &  &  \\ 
  CLAIMSx & Initial Claims &   5 & x & x & EXCAUSx & Canada / U.S. Foreign Exchange Rate &   5 &  &  \\ 
  PAYEMS & All Employees:  Total nonfarm &   5 & x & x & WPSFD49207 & PPI: Finished Goods &   6 &  &  \\ 
  USGOOD & All Employees:  Goods-Producing Industries &   5 &  &  & WPSFD49502 & PPI: Finished Consumer Goods &   6 &  &  \\ 
  CES1021000001 & All Employees:  Mining and Logging:  Mining &   5 &  &  & WPSID61 & PPI: Intermediate Materials &   6 &  &  \\ 
  USCONS & All Employees:  Construction &   5 &  &  & WPSID62 & PPI: Crude Materials &   6 &  &  \\ 
  MANEMP & All Employees:  Manufacturing &   5 &  &  & OILPRICEx & Crude Oil, spliced WTI and Cushing &   6 &  & x \\ 
  DMANEMP & All Employees:  Durable goods &   5 &  &  & PPICMM & PPI: Metals and metal products: &   6 &  &  \\ 
  NDMANEMP & All Employees:  Nondurable goods &   5 &  &  & CPIAUCSL & CPI : All Items &   6 & x & x \\ 
  SRVPRD & All Employees:  Service-Providing Industries &   5 &  &  & CPIAPPSL & CPI : Apparel &   6 &  &  \\ 
  USTPU & All Employees:  Trade, Transportation \& Utilities &   5 &  &  & CPITRNSL & CPI : Transportation &   6 &  &  \\ 
  USWTRADE & All Employees:  Wholesale Trade &   5 &  &  & CPIMEDSL & CPI : Medical Care &   6 &  &  \\ 
  USTRADE & All Employees:  Retail Trade &   5 &  &  & CUSR0000SAC & CPI : Commodities &   6 &  &  \\ 
  USFIRE & All Employees:  Financial Activities &   5 &  &  & CUSR0000SAD & CPI : Durables &   6 &  &  \\ 
  USGOVT & All Employees:  Government &   5 &  &  & CUSR0000SAS & CPI : Services &   6 &  &  \\ 
  CES0600000007 & Avg Weekly Hours :  Goods-Producing &   1 & x & x & CPIULFSL & CPI : All Items Less Food &   6 &  &  \\ 
  AWOTMAN & Avg Weekly Overtime Hours :  Manufacturing &   2 &  &  & CUSR0000SA0L2 & CPI : All items less shelter &   6 &  &  \\ 
  AWHMAN & Avg Weekly Hours :  Manufacturing &   1 &  & x & CUSR0000SA0L5 & CPI : All items less medical care &   6 &  &  \\ 
  HOUST & Housing Starts:  Total New Privately Owned &   4 & x & x & PCEPI & Personal Cons.  Expend.:  Chain Index &   6 &  & x \\ 
  HOUSTNE & Housing Starts, Northeast &   4 &  &  & DDURRG3M086SBEA & Personal Cons.  Exp:  Durable goods &   6 &  &  \\ 
  HOUSTMW & Housing Starts, Midwest &   4 &  &  & DNDGRG3M086SBEA & Personal Cons.  Exp:  Nondurable goods &   6 &  &  \\ 
  HOUSTS & Housing Starts, South &   4 &  &  & DSERRG3M086SBEA & Personal Cons.  Exp:  Services &   6 &  &  \\ 
  HOUSTW & Housing Starts, West &   4 &  &  & CES0600000008 & Avg Hourly Earnings :  Goods-Producing &   6 &  & x \\ 
  PERMIT & New Private Housing Permits (SAAR) &   4 &  & x & CES2000000008 & Avg Hourly Earnings :  Construction &   6 &  &  \\ 
  PERMITNE & New Private Housing Permits, Northeast (SAAR) &   4 &  &  & CES3000000008 & Avg Hourly Earnings :  Manufacturing &   6 &  &  \\ 
  PERMITMW & New Private Housing Permits, Midwest (SAAR) &   4 &  &  & DTCOLNVHFNM & Consumer Motor Vehicle Loans Outstanding &   6 &  &  \\ 
  PERMITS & New Private Housing Permits, South (SAAR) &   4 &  &  & DTCTHFNM & Total Consumer Loans and Leases Outstanding &   6 &  &  \\ 
  PERMITW & New Private Housing Permits, West (SAAR) &   4 &  &  & INVEST & Securities in Bank Credit at All Commercial Banks &   6 &  &  \\ 
  AMDMNOx & New Orders for Durable Goods &   5 &  &  & VIXCLSx & VIX &   1 &  &  \\ 
   \bottomrule
\end{tabular}}
\end{tabular}
\begin{tablenotes}[para,flushleft]
\footnotesize \textit{Notes}: The column \textit{Trans.} denotes the following data transformation for a series x: (1) no transformation; (2) $\Delta x_t$; (3) $\Delta^2 x_t$; (4) $log(x_t)$; (5) $\Delta log(x_t)$; (6) $\Delta^2 log(x_t)$; (7) $\Delta(x_t/x_{t-1}-1)$; (8) $100 \times \left((x_t/x_{t-1})^4 - 1\right)$. The columns \textit{S} and \textit{M} indicate that the variable is included in small or medium information set, respectively.
\end{tablenotes}
\end{threeparttable}
\end{table}

\clearpage
\subsection{Additional empirical results}
\begin{table}[!ht]
\includegraphics[width=\textwidth]{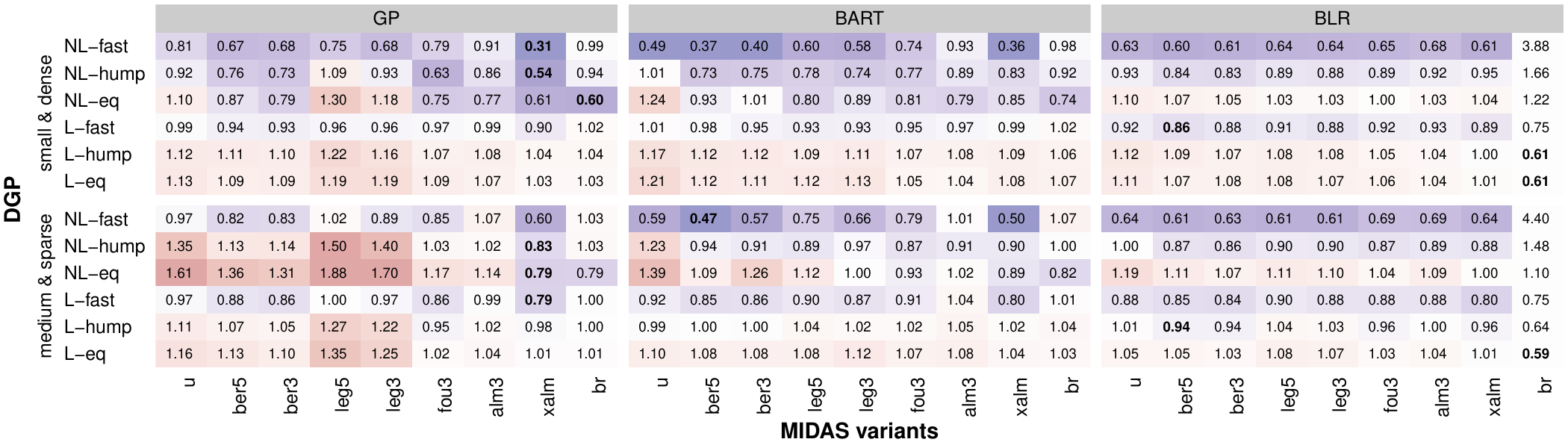}
\caption{Comparison of predictive (nowcast) losses using MAE against the \texttt{BLR}-\texttt{br} benchmark across different model specifications and data generating processes (DGPs). \textit{Notes}: Each row represents a different DGP, columns denote different model specifications. DGPs vary according to the functional form --- \texttt{NL} (nonlinear) vs. \texttt{L} (linear), according to the weighting scheme --- fast-decaying (\texttt{fast}), hump-shaped (\texttt{hump}), vs. equal (\texttt{eq}) weights, and according to model complexity --- medium ($K = 25$) and sparse ($80\%$ sparsity) vs. small ($K = 10$) and (relatively) dense ($50\%$ sparsity). The best-performing specification for each DGP is highlighted in bold. We simulate $R = 50$ replications for each DGP.}
\label{tab:MAEsynthetic}
\end{table}

\begin{table}[ht]
    \centering
    \begin{subfigure}[t]{\textwidth}
        \caption{\texttt{GDPC1}}
        \includegraphics[width=\textwidth]{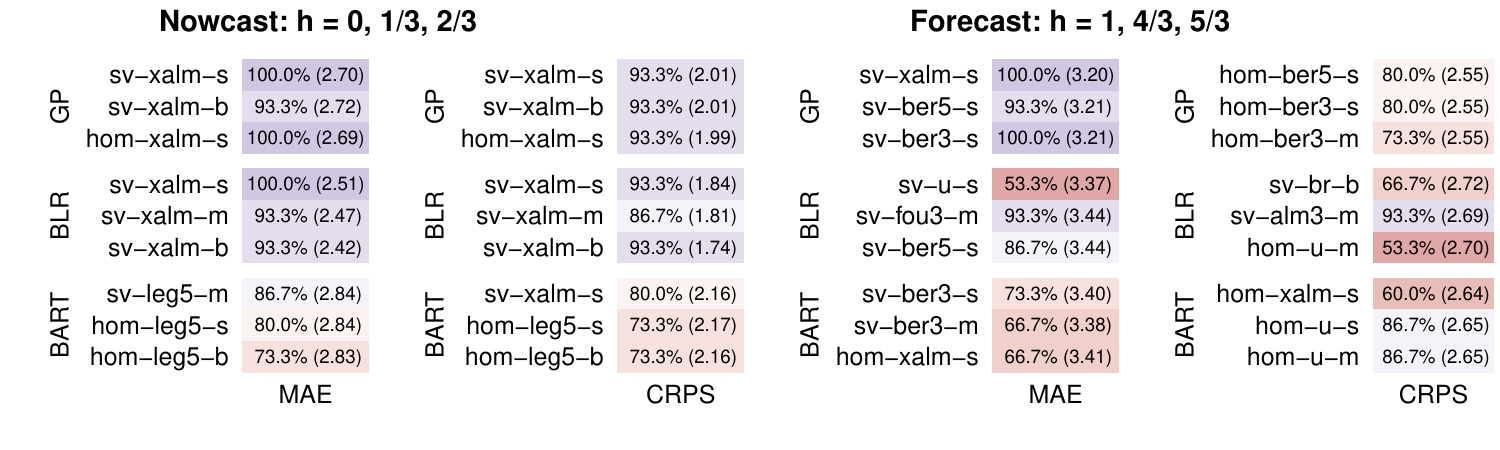}
    \end{subfigure}
    \begin{subfigure}[t]{\textwidth}
        \caption{\texttt{GDPCTPI}}
        \includegraphics[width=\textwidth]{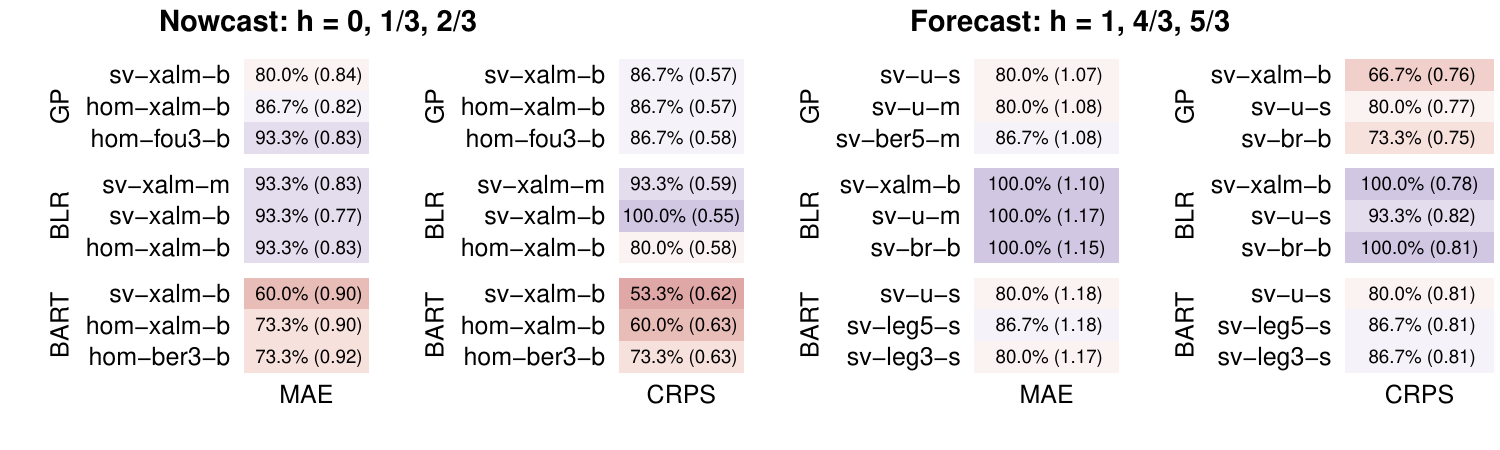}
    \end{subfigure}
    
    \caption{Summary of model confidence sets across horizons and subsamples. \textit{Notes}: Percentages indicate how often the respective specification was among the superior model set, the number in parentheses are averaged losses.}
    \label{fig:mcs2}
\end{table}

\begin{figure}[ht]
    \begin{subfigure}{\textwidth}
        \centering\caption{\texttt{GDPC1}}
        \includegraphics[width=0.49\textwidth]{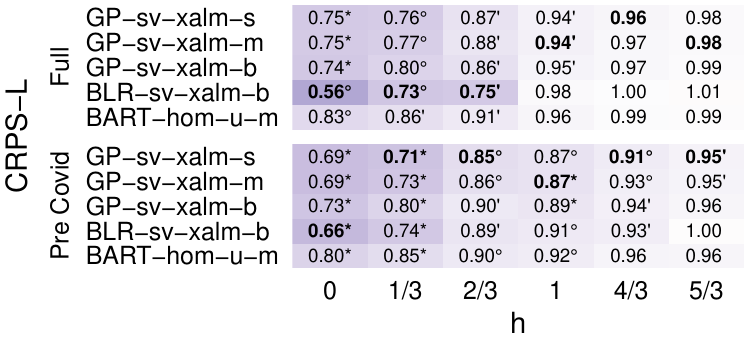}
        \includegraphics[width=0.49\textwidth]{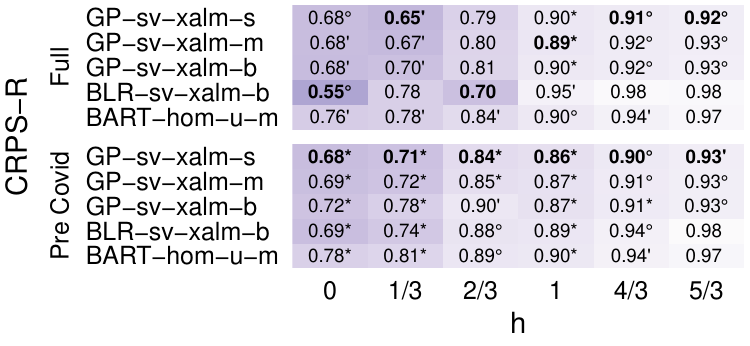}
    \end{subfigure}
    \begin{subfigure}{\textwidth}
        \centering\caption{\texttt{GDPCTPI}}
        \includegraphics[width=0.49\textwidth]{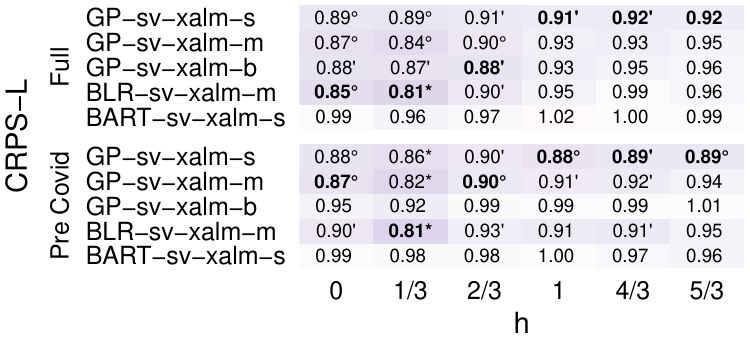}
        \includegraphics[width=0.49\textwidth]{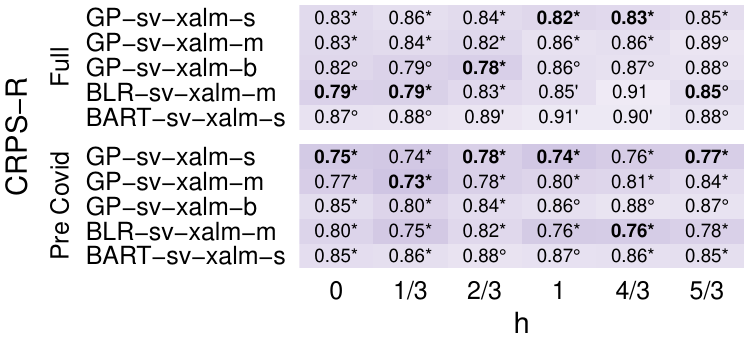}
    \end{subfigure}
    \caption{Predictive losses for selected specifications relative to the benchmark AR($P_L$)-model. \textit{Notes}: Levels of statistical significance $\{$5, 1, 0.1$\}$\% for a DM-test of equal predictive accuracy are indicated with $\{',^\circ,^\ast\}$. Best-performing specification by horizon in bold.}
    \label{tab:crps_tails_selected}
\end{figure}

\begin{table}[ht]
    \includegraphics[width=0.49\textwidth]{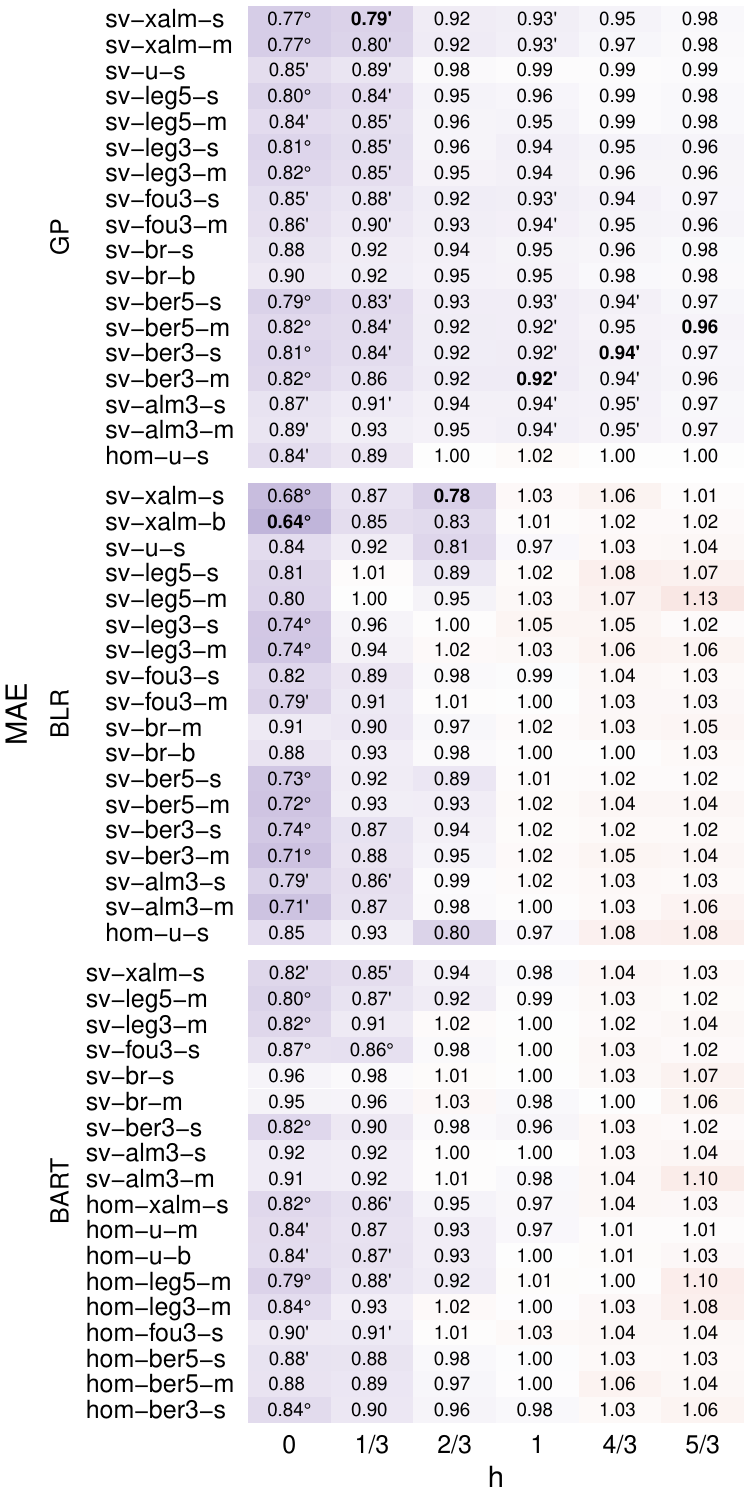}
    \includegraphics[width=0.49\textwidth]{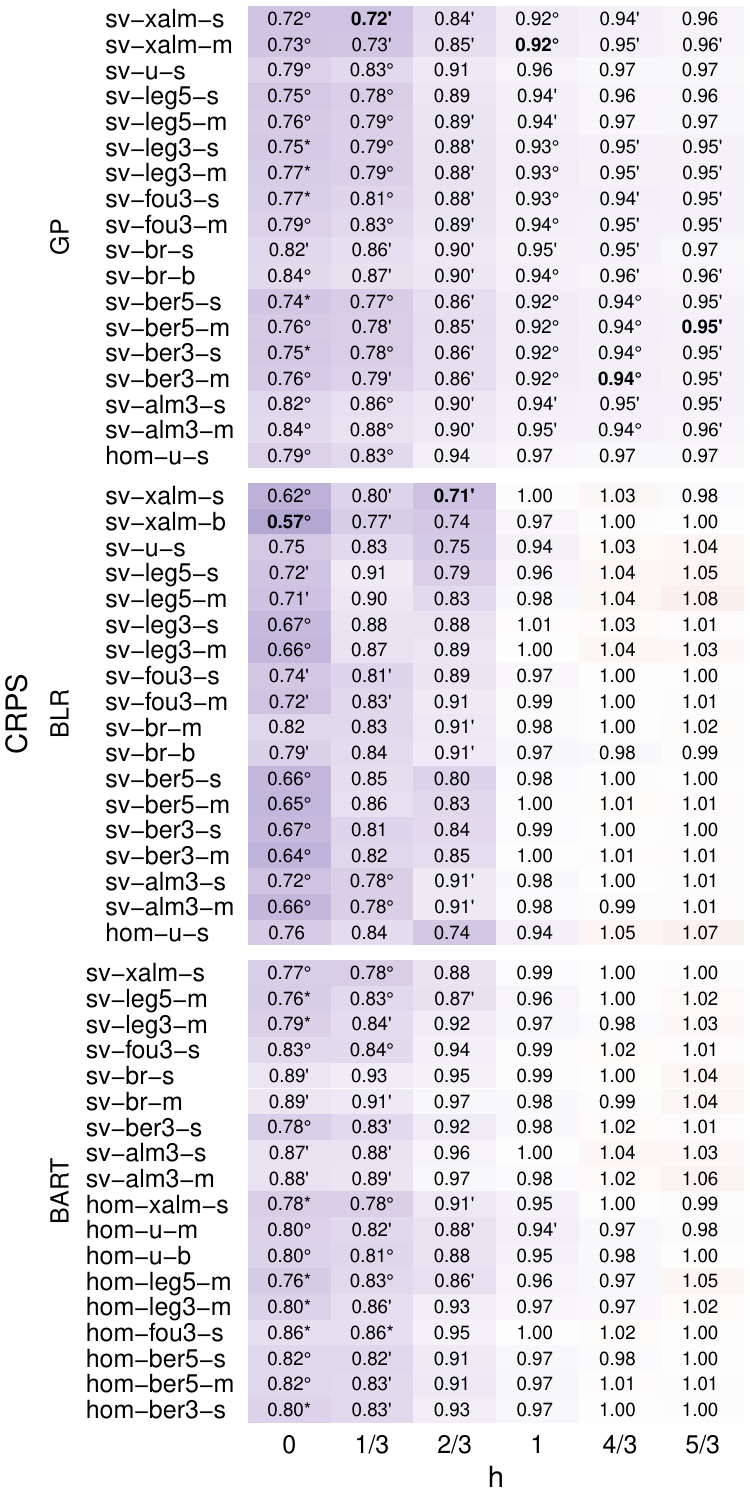}
    
    \caption{Predictive losses over the full holdout for GDPC1, best specification for each conditional mean and MIDAS-type combination, relative to the benchmark AR($P_L$)-model. \textit{Notes}: Levels of statistical significance $\{$5, 1, 0.1$\}$\% for a DM-test of equal predictive accuracy are indicated with $\{',^\circ,^\ast\}$. Best-performing specification by column in bold.}
    \label{tab:crps_gdpc1_appendix}
\end{table}

\begin{table}[ht]
    \includegraphics[width=0.49\textwidth]{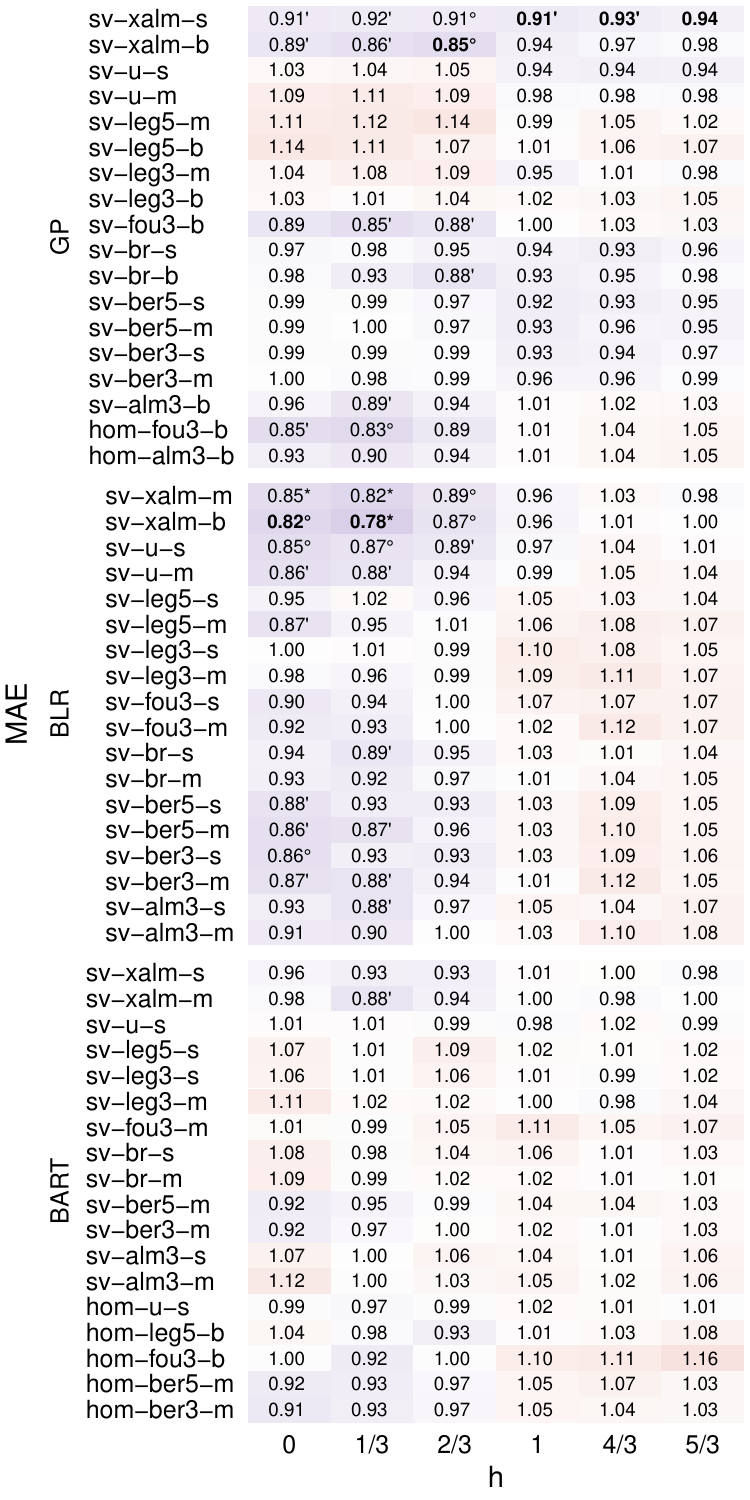}
    \includegraphics[width=0.49\textwidth]{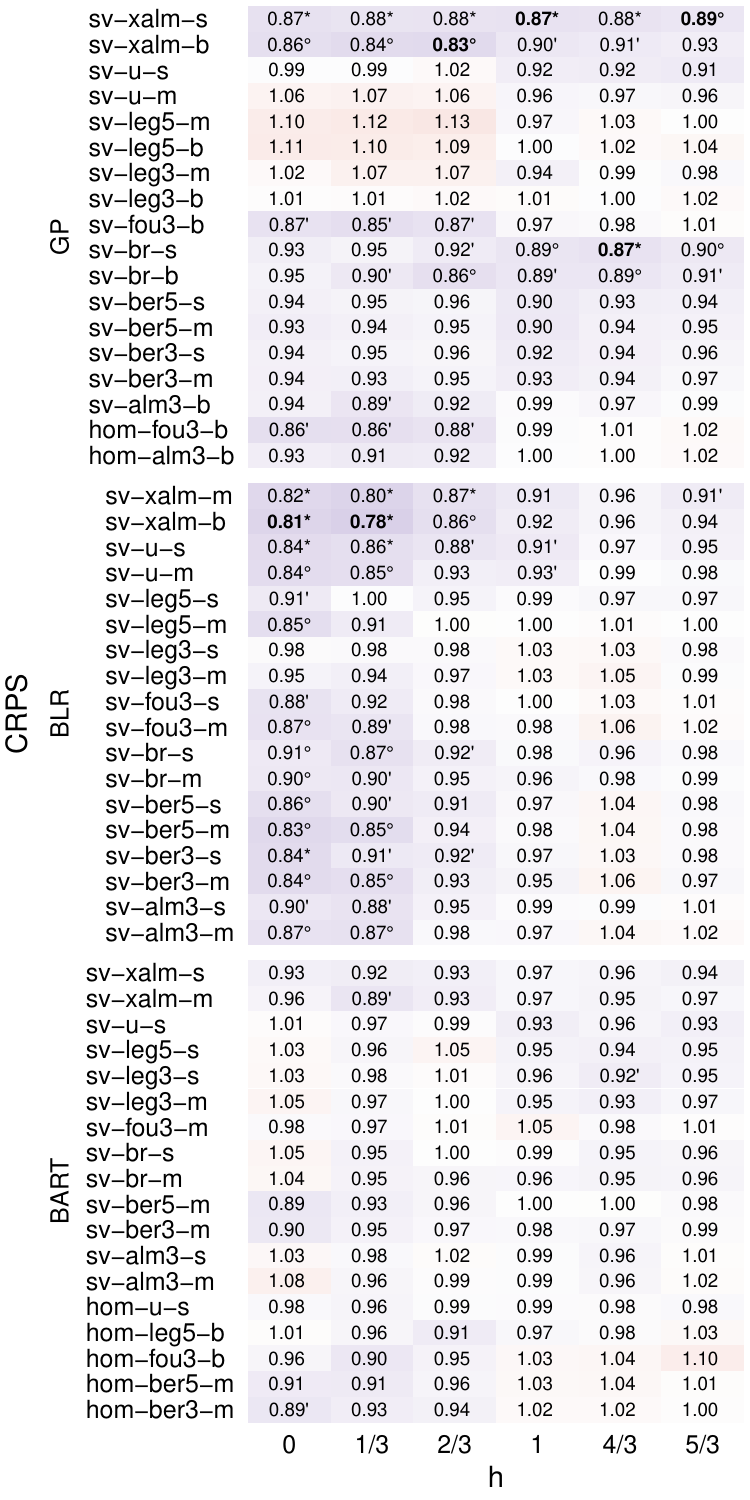}
    
    \caption{Predictive losses over the full holdout for GDPCTPI, best specification for each conditional mean and MIDAS-type combination, relative to the benchmark AR($P_L$)-model. \textit{Notes}: Levels of statistical significance $\{$5, 1, 0.1$\}$\% for a DM-test of equal predictive accuracy are indicated with $\{',^\circ,^\ast\}$. Best-performing specification by column in bold.}
    \label{tab:crps_gdpctpi_appendix}
\end{table}

\clearpage
\subsection{Measuring variable importance}
\begin{table}[ht]
    \includegraphics[width=\textwidth]{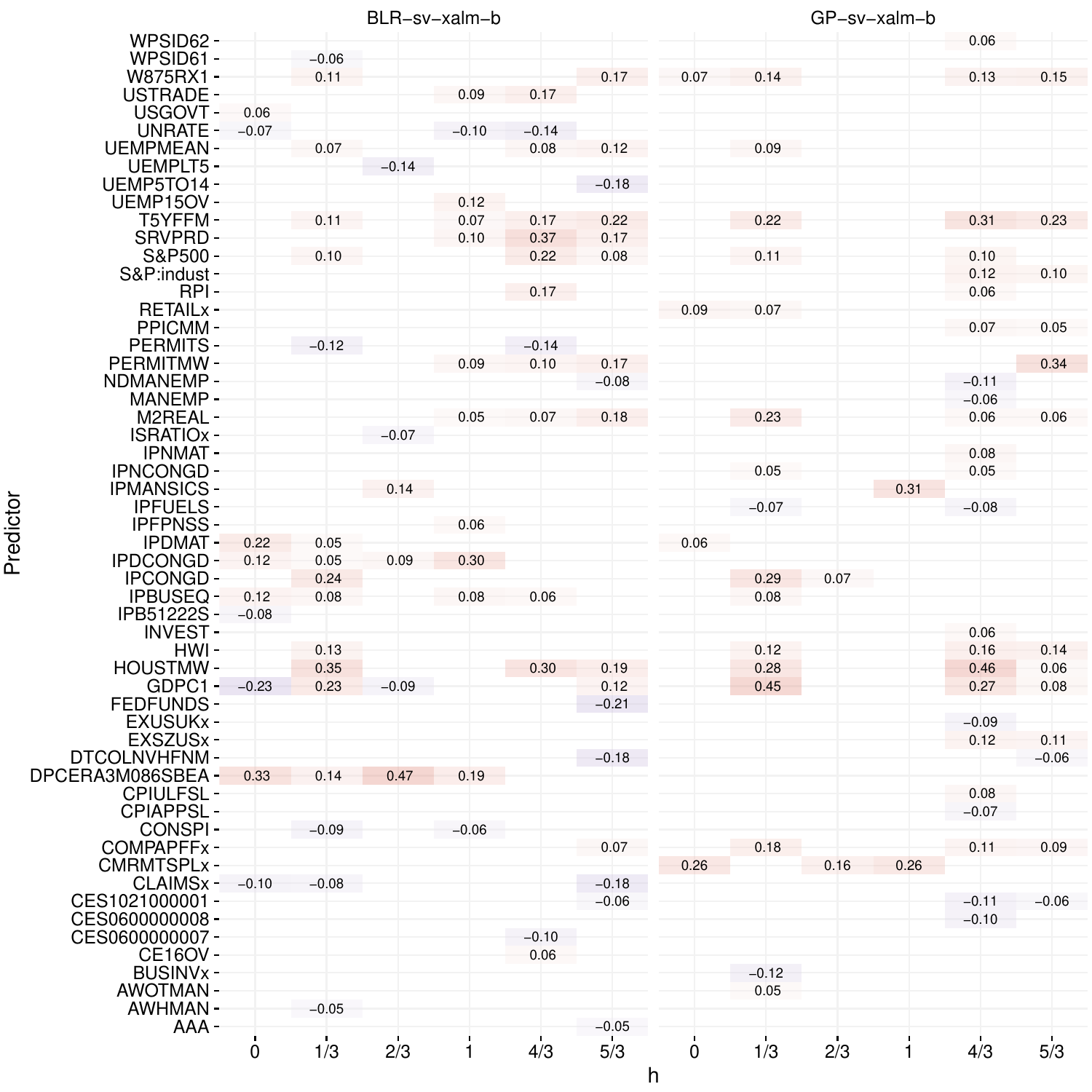}
    \caption{Most important predictors for GDPC1 over the holdout for selected model specifications across predictive horizons. \textit{Notes}: Values are standardized coefficients obtained from a \textsc{Lasso} regression.}
    \label{tab:varimp_gdpc1}
\end{table}

\begin{table}[ht]
    \includegraphics[width=\textwidth]{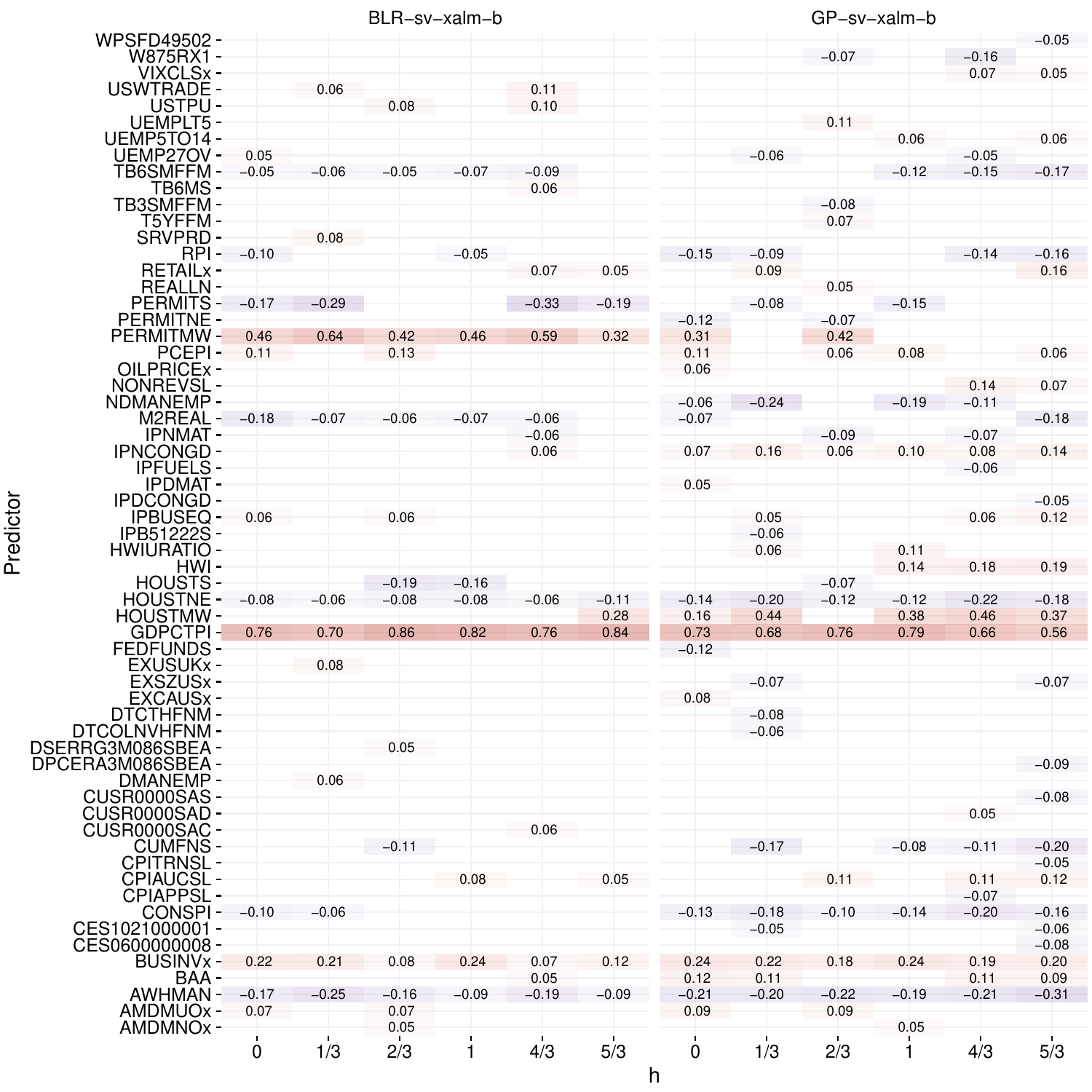}
    \caption{Most important predictors for GDPCTPI over the holdout for selected model specifications across predictive horizons. \textit{Notes}: Values are standardized coefficients obtained from a \textsc{Lasso} regression.}
    \label{tab:varimp_gdpctpi}
\end{table}

\end{appendices}
\end{document}